\documentclass[pdftex,twocolumn,epjc3]{svjour3}          

\RequirePackage[T1]{fontenc}

\smartqed  

\RequirePackage{graphicx}
\RequirePackage{mathptmx}      
\RequirePackage{flushend}
\RequirePackage[numbers,sort&compress]{natbib}
\RequirePackage[colorlinks,citecolor=blue,urlcolor=blue,linkcolor=blue]{hyperref}

\newcommand{\orcid}[1]{\href{https://orcid.org/#1}{\,\includegraphics[width=8px]{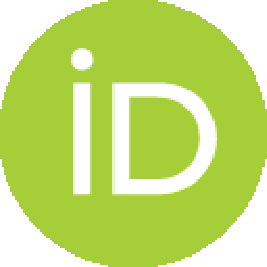}}}

\usepackage{amsmath,amssymb}

\journalname{Eur. Phys. J. C}

\begin{document}

\title{Analytical Gaussian Process Cosmography: Unveiling Insights into Matter-Energy Density Parameter at Present}


\author{Bikash R. Dinda\orcid{0000-0001-5432-667X}\thanksref{e1,addr1,addr2}
}

\thankstext{e1}{e-mail: bikashd18@gmail.com}

\institute{Department of Physical Sciences, Indian Institute of Science Education and Research Kolkata, Mohanpur, Nadia, West Bengal 741246, India.\label{addr1}
         \and
Department of Physics $\&$ Astronomy, University of the Western Cape, Cape Town, 7535, South Africa.\label{addr2}
}

\date{Received: date / Accepted: date}

\maketitle

\begin{abstract}
In this study, we introduce a novel analytical Gaussian Process (GP) cosmography methodology, leveraging the differentiable properties of GPs to derive key cosmological quantities analytically. Our approach combines cosmic chronometer (CC) Hubble parameter data with growth rate (f) observations to constrain the $\Omega_{\rm m0}$ parameter, offering insights into the underlying dynamics of the Universe. By formulating a consistency relation independent of specific cosmological models, we analyze under a flat FLRW metric and first-order Newtonian perturbation theory framework. Our analytical approach simplifies the process of Gaussian Process regression (GPR), providing a more efficient means of handling large datasets while offering deeper interpretability of results. We demonstrate the effectiveness of our methodology by deriving precise constraints on $\Omega_{\rm m0}h^2$, revealing $\Omega_{\rm m0}h^2=0.139\pm0.017$. Moreover, leveraging $H_0$ observations, we further constrain $\Omega_{\rm m0}$, uncovering an inverse correlation between mean $H_0$ and $\Omega_{\rm m0}$. Our investigation offers a proof of concept for analytical GP cosmography, highlighting the advantages of analytical methods in cosmological parameter estimation.
\end{abstract}

\section{Introduction}

In 1998, observations of type Ia supernovae provided compelling evidence for the current accelerated expansion of the Universe, marking the onset of late-time cosmic acceleration \cite{SupernovaCosmologyProject:1997zqe,SupernovaSearchTeam:1998fmf,SupernovaCosmologyProject:1998vns,2011NatPh...7Q.833W,Linden2009CosmologicalPE,Camarena:2019rmj,Pan-STARRS1:2017jku,Camlibel:2020xbn}. Subsequent observations, including those of the cosmic microwave background (CMB) \cite{Planck:2013pxb,Planck:2015fie,Planck:2018vyg}, cosmic chronometers (CC) measuring the Hubble parameter \citep{Jimenez:2001gg,Pinho:2018unz,Cao:2023eja}, and baryon acoustic oscillations (BAO) \citep{BOSS:2016wmc,eBOSS:2020yzd,Hou:2020rse}, have independently confirmed this accelerated expansion. The prevailing explanation for this phenomenon revolves around two main classes of models: one posits the existence of dark energy, a component with a large negative pressure driving the acceleration \cite{Peebles:2002gy,Copeland:2006wr,Yoo:2012ug,Lonappan:2017lzt,Dinda:2017swh,Dinda:2018uwm}; the other considers modifications to the general theory of relativity on cosmological scales \cite{Clifton:2011jh,Koyama:2015vza,Tsujikawa:2010zza,Joyce:2016vqv,Dinda:2017lpz,Dinda:2018eyt,Zhang:2020qkd,Dinda:2022ixi,Bassi:2023vaq,Nojiri:2010wj,Nojiri:2017ncd,Bamba:2012cp,Lee:2022cyh}. Among these models, the $\Lambda$CDM model stands out as the most popular and successful \citep{Carroll:2000fy}. In $\Lambda$CDM, the cosmological constant is identified as the leading candidate for dark energy, providing a robust framework for understanding the observed cosmic acceleration.

Despite the considerable success of the $\Lambda$CDM model, it grapples with both theoretical and observational challenges. Theoretical concerns include issues of fine-tuning and the cosmic coincidence problem \citep{Zlatev:1998tr,Sahni:1999gb,Velten:2014nra,Malquarti:2003hn}. On the observational front, the model exhibits discrepancies in derived quantities such as $H_0$ (the present Hubble parameter) \citep{DiValentino:2021izs,Krishnan:2021dyb,Vagnozzi:2019ezj,Dinda:2021ffa} and $\sigma_8$ parameters, particularly between early-time observations like the CMB and late-time measurements such as local determinations of $H_0$ and cosmic shear observations of $\sigma_8$ \citep{DiValentino:2020vvd,Abdalla:2022yfr,Douspis:2018xlj,Bhattacharyya:2018fwb,Bargiacchi:2023rfd}. These inconsistencies have spurred investigations beyond the $\Lambda$CDM paradigm.

Efforts to address these challenges have led to explorations of alternative models, including dynamical dark energy models \cite{Lonappan:2017lzt} and early dark energy models \cite{Abdalla:2022yfr}. While these models have shown success to a certain extent, concrete solutions to the identified problems remain elusive. Consequently, in recent times, there has been a shift towards model-independent and non-parametric approaches in the analysis \cite{Haridasu:2018gqm,Bernardo:2021mfs,Wei:2020suh,Naik:2023yhl,Capozziello:2021xjw,Alfano:2023evg,Capozziello:2014zda,Dinda:2022vmb,Dinda:2022jih}. These approaches aim to explore and study various cosmological observables, such as the Hubble parameter \cite{Bernardo:2021mfs,Wei:2020suh} and the deceleration parameter \cite{Naik:2023yhl,Capozziello:2021xjw,Alfano:2023evg,Capozziello:2014zda}, without being constrained by specific theoretical frameworks.

In the literature, certain parametric approaches are colloquially labeled as 'model-independent' because these parametrizations ostensibly avoid explicit dependence on specific cosmological models \cite{Dinda:2021ffa,Dinda:2019mev}. An illustrative example is the cosmographic approach, which characterizes the expansion of the universe through distinct redshift or time derivatives of the scale factor or Hubble parameters \cite{Dinda:2019mev}. However, it is essential to discern that these parametrizations, though not directly rooted in particular cosmological models, can themselves be considered models.

In this study, when we refer to model-independent analysis, we explicitly denote an approach that is truly free from reliance on any particular theoretical model or parametrization. While acknowledging the necessity of foundational concepts such as the existence of standard candles, our pursuit of model independence extends to a genuine absence of reliance on any cosmological model governing background expansion or the evolution of inhomogeneities, whether at the first order or even at higher-order perturbations \cite{Ruiz-Zapatero:2022zpx,Avila:2022xad,LHuillier:2017ani,Lee:2013sya,Holanda:2019sod}.

To adapt the model-independent methodology, we use posterior Gaussian process regression (GPR) analysis \citep{williams1995gaussian,GpRasWil,Seikel_2012,PhysRevD.85.123530,Hwang:2022hla,Keeley:2020aym,Dinda:2022vmb,Dinda:2022jih,Perenon:2022fgw,Perenon:2021uom,OColgain:2021pyh,Banerjee:2023evd,Banerjee:2023rvg,Mukherjee:2022ujw,Mukherjee:2020vkx,Mukherjee:2020ytg,Li:2019nux,Ruiz-Zapatero:2022zpx,Ruiz-Zapatero:2022xbv}. GPR is a powerful statistical technique used in various fields, including cosmology, to model complex data relationships. In the context of cosmography, GPR enables the reconstruction of cosmological observables, such as the Hubble parameter and the growth rate of cosmic structures, from observational data. Traditionally, GPR involves numerically sampling the Gaussian processes, which can be computationally intensive, especially with large datasets. However, with advancements in analytical methods, such as the analytical Gaussian process, we can now perform GPR more efficiently \citep{Hwang:2022hla}.

The analytical approach offers several advantages over numerical methods. Firstly, it simplifies the computational burden by avoiding the need for extensive numerical sampling. This makes it particularly advantageous for handling large cosmological datasets, where computational efficiency is crucial. Secondly, analytical GPR provides insights into the underlying data relationships in a more interpretable manner, facilitating a deeper understanding of cosmological phenomena \citep{Hwang:2022hla,PhysRevD.85.123530,Seikel_2012}.

Despite its benefits, analytical GPR also imposes constraints on the data that can be effectively fitted. These constraints arise from the assumptions inherent in the analytical framework and must be carefully considered to ensure the reliability of the results. Therefore, while analytical GPR offers a promising avenue for cosmography, it requires careful validation and verification to ensure its applicability to specific datasets \citep{Li:2019nux,Ruiz-Zapatero:2022zpx,Ruiz-Zapatero:2022xbv}.

Overall, the integration of analytical GPR into cosmological methodologies represents a significant advancement in our ability to extract meaningful insights from observational data. By leveraging the simplicity and efficiency of analytical techniques, we can enhance the accuracy and robustness of cosmological parameter estimation, ultimately advancing our understanding of the universe's fundamental properties and evolution \citep{Hwang:2022hla,PhysRevD.85.123530,Seikel_2012}.

Cosmological observations offer valuable insights, shedding light not just on the late-time cosmic acceleration but also on the current composition of the Universe. Currently, dark energy constitutes approximately 70$\%$ of the total energy budget, with total matter contributing around 30$\%$. This distribution implies a present matter energy density parameter, denoted as $\Omega_{\rm m0}$, of approximately 0.3 \cite{Planck:2013pxb,Planck:2015fie,Planck:2018vyg}.

However, asserting $\Omega_{\rm m0} \approx 0.3$ lacks the straightforwardness of the evidence for late-time cosmic acceleration in terms of model-independent analysis. The dominance of dark energy in the late stages of the Universe's evolution sufficiently explains the observed acceleration. Determining cosmological quantities such as the deceleration parameter ($q$) or the Hubble parameter ($H$) in a model-independent manner can be achieved through a single type of cosmological observation or by cross-calibrating different datasets \citep{Sakr:2023hrl,Ruiz-Zapatero:2022zpx}.

For instance, $H$ can be derived exclusively from cosmic chronometer observations, while $q$ can be obtained from the derivative of the Hubble parameter data, provided the derivative is computed using model-independent techniques \cite{Dinda:2022vmb,Dinda:2022jih}. However, many background cosmological observations predominantly involve the Hubble parameter or cosmological distances, like the luminosity distance. Relying solely on a single type of observation or calibration between them is insufficient to determine the precise value of $\Omega_{\rm m0}$. This limitation arises because neither the Hubble parameter nor cosmological distances trace the individual energy budget of each constituent in the Universe.

Cosmological observations closely associated with the growth rate, denoted as $f$, of matter inhomogeneities, play a crucial role in determining the value of the $\Omega_{\rm m0}$ parameter \cite{Ruiz-Zapatero:2022zpx,Avila:2022xad,LHuillier:2017ani,Lee:2013sya}. This study aims to integrate these observations with background cosmological data, particularly insights from cosmic chronometers. The goal is to ascertain $\Omega_{\rm m0}$ in a manner independent of specific models. This approach becomes feasible as we will demonstrate that the equation governing the evolution of inhomogeneity growth explicitly features the $\Omega_{\rm m0}$ parameter.

The paper is structured as follows: Section~\ref{sec-background} outlines the redshift evolution of the matter-energy density parameter. Section~\ref{sec-perturbations} derives the first-order perturbation theory equation for the matter growth rate. Section~\ref{sec-error_propagation} details the expression for uncertainty propagation. Section~\ref{sec-data} briefly discusses cosmic chronometers and growth rate data. Section~\ref{sec-method} delves into the methodology of Gaussian process regression analysis. Section~\ref{sec-result} presents the constraints on the present matter energy density parameter. Finally, Section~\ref{sec-conclusion} concludes the study.

\section{Matter energy density parameter}
\label{sec-background}

We posit that at late times, the Universe is predominantly governed by matter and dark energy. In this context, 'matter' encompasses both cold dark matter and baryons. Additionally, we assume that there is no interaction between dark energy and matter. With these considerations and under the assumption of a flat Friedmann-Lemaître-Robertson-Walker (FLRW) metric for the background expansion of the Universe, the evolution of the background matter energy density, denoted as $\bar{\rho}_m$, is derived as follows

\begin{equation}
\bar{\rho}_m = \bar{\rho}_{\rm m0} (1+z)^3,
\label{eq:rho_m_bar}
\end{equation}

\noindent
where $\bar{\rho}_{\rm m0}$ represents the current value of the matter-energy density, and $z$ denotes the redshift. Using the above equation, the evolution of the matter-energy density parameter is expressed as \citep{Perenon:2022fgw}

\begin{equation}
\Omega_m = \frac{\bar{\rho}_m}{3M_{\rm pl}^2H^2} = \frac{ \bar{\rho}_{\rm m0} H_0^2 (1+z)^3 }{3M_{\rm pl}^2H_0^2 H^2} = \frac{\Omega_{\rm m0}H_0^2(1+z)^3}{H^2},
\label{eq:Omega_m}
\end{equation}

\noindent
where $H$ represents the Hubble parameter, and $H_0$ is its current value. The reduced Planck mass is denoted as $M_{\rm pl}$, while $\Omega_{\rm m0}$ signifies the present value of the matter-energy density parameter.

\section{Growth of matter inhomogeneties}
\label{sec-perturbations}

In the sub-Hubble limit and within the linear regime, employing first-order linear Newtonian perturbation theory allows us to investigate the evolution of perturbations in the Universe. In this context, the differential equation governing the growth of matter inhomogeneity, denoted as $\delta_m$, is presented as \citep{Dinda:2018uwm, Dinda:2023mad, Perenon:2022fgw, Dinda:2017swh, Dinda:2018ojk}:

\begin{equation}
\ddot{\delta}_m+2H\dot{\delta}_m-4\pi G \bar{\rho}_m \delta_m=0,
\label{eq:basic_matter_inhomogeneity}
\end{equation}

\noindent
where the overhead dot and double-dot signify first and second-order differentiations with respect to cosmic time $t$, and $G$ represents the Newtonian gravitational constant. The differential equation above yields two solutions for $\delta_m$: one associated with the growing mode and the other with the decaying mode. Our focus is on the growing mode solution, denoted as $D_+$. This $D_+$ follows the same differential equation as in Eq.~\eqref{eq:basic_matter_inhomogeneity}.

We use the notation $Q'=\frac{dQ}{dz}$ and $Q''=\frac{d^2Q}{dz^2}$, where primes and double primes represent first and second-order differentiations with respect to redshift $z$. Utilizing the relations $\dot{D}_+=-(1+z)H D'_+$, $\ddot{D}_+=(1+z)^2H^2 \left[ D''_+ + (1+\frac{1+z}{H} H')D'_+ \right] $, and $8\pi G=M_{\rm pl}^{-2}$, we express the differential equation for $D_+$ as

\begin{equation}
(1+z)^2 D''_+ + (1+z) \left( \frac{1+z}{H} H'-1 \right) D'_+ - \frac{3}{2} \Omega_m D_+ = 0.
\label{eq:growing_mode}
\end{equation}

In the context of cosmic structure formation, a crucial quantity is the logarithmic growth rate $f$, defined as

\begin{equation}
f = \dfrac{d \ln D_+}{d \ln a} = - \frac{1+z}{D_+} D'_+,
\label{eq:defn_f}
\end{equation}

\noindent
where $a$ represents the cosmic scale factor, and its relationship with redshift is given by $1+z=a^{-1}$. The equation above can be reformulated as $D'_+ =-\frac{D_+}{1+z} f$. Differentiating this equation yields $D''_+ = -\frac{D_+}{(1+z)^2} \left[ (1+z)f'-f^2-f \right]$. Substituting these relations into Eq.~\eqref{eq:growing_mode}, we obtain a differential equation for $f$ given as \cite{Ruiz-Zapatero:2022zpx}

\begin{equation}
(1+z) f' - f^2 + \left( \frac{1+z}{H} H'-2 \right) f + \frac{3}{2} \Omega_m = 0.
\label{eq:diff_eqn_f}
\end{equation}

\noindent
This equation can be rearranged to explicitly express $\Omega_m$ in terms of $f$, $f'$, and other relevant quantities, as given below

\begin{equation}
\Omega_m = \frac{2}{3H} \left[ \left( 2H-(1+z)H' \right) f + Hf^2 -(1+z)Hf' \right].
\label{eq:Omega_m_from_f}
\end{equation}

By equating Eqs.~\eqref{eq:Omega_m} and~\eqref{eq:Omega_m_from_f}, an expression for $\Omega_{\rm m0} H_0^2$ emerges as

\begin{equation}
\Omega_{\rm m0} H_0^2 = \frac{2H \left[ (2+f)Hf-(1+z)(H'f+Hf') \right] }{3(1+z)^3}.
\label{eq:Omega_m0_from_H_and_f}
\end{equation}

\noindent
From this equation, one can estimate the value of $\Omega_{\rm m0} H_0^2$ by knowing the values of $H$, $H'$, $f$, and $f'$ at a specific redshift $z$.

\section{Propagation of errors}
\label{sec-error_propagation}

Let us define

\begin{equation}
W_{\rm m0} = \Omega_{\rm m0} H_0^2.
\label{eq:Wm0_from_Omega_m0_H0sqr}
\end{equation}

\noindent
The accuracy of the estimated $W_{\rm m0}$ using Eq.~\eqref{eq:Omega_m0_from_H_and_f} is contingent upon the uncertainties in $H$, $H'$, $f$, and $f'$. To quantify this uncertainty, we employ the propagation of errors, expressed as

\begin{eqnarray}
\text{Var}[W_{\rm m0}] &=& \left( \frac{\partial W_{\rm m0}}{\partial H} \right)^2 \text{Var}[H] + \left( \frac{\partial W_{\rm m0}}{\partial H'} \right)^2 \text{Var}[H'] \nonumber\\
&& + 2 \left( \frac{\partial W_{\rm m0}}{\partial H} \right) \left(  \frac{\partial W_{\rm m0}}{\partial H'} \right) \text{Cov}[H,H'] \nonumber\\
&& + \left( \frac{\partial W_{\rm m0}}{\partial f} \right)^2 \text{Var}[f] + \left( \frac{\partial W_{\rm m0}}{\partial f'} \right)^2 \text{Var}[f'] \nonumber\\
&& + 2 \left( \frac{\partial W_{\rm m0}}{\partial f} \right) \left(  \frac{\partial W_{\rm m0}}{\partial f'} \right) \text{Cov}[f,f'],
\label{eq:var_Wm0}
\end{eqnarray}

\noindent
where $\text{Var}[Q]$ denotes the variance in the quantity $Q$, and $\text{Cov}[P,Q]$ represents the covariance between two quantities, $P$ and $Q$. And we have

\begin{eqnarray}
\frac{\partial W_{\rm m0}}{\partial H} &=& \frac{4 f^2 H-4 H (z+1) f'-2 f (z+1) H'+8 f H}{3 (z+1)^3},
\label{eq:dWm0dH} \\
\frac{\partial W_{\rm m0}}{\partial H'} &=& -\frac{2 f H}{3 (z+1)^2},
\label{eq:dWm0dHprime} \\
\frac{\partial W_{\rm m0}}{\partial f} &=& \frac{2 H \left(2 (f+1) H-(z+1) H'\right)}{3 (z+1)^3},
\label{eq:dWm0df} \\
\frac{\partial W_{\rm m0}}{\partial f'} &=& -\frac{2 H^2}{3 (z+1)^2}.
\label{eq:dWm0dfprime}
\end{eqnarray}

It's important to note that in Eq.~\eqref{eq:var_Wm0}, we specifically account for the covariance between $H$ and $H'$, as well as the covariance between $f$ and $f'$, excluding other pairs. The rationale behind this selective consideration will be discussed in subsequent sections.

\section{Observational data}
\label{sec-data}

In our investigation, we incorporate data from cosmic chronometer (CC) observations, comprising a comprehensive set of 32 Hubble parameter measurements distributed across a range of redshift values ($0.07 \leq z \leq 1.965$). This dataset, as meticulously detailed in \cite{Cao:2023eja}, plays a pivotal role in unraveling the intricacies of cosmic evolution. It's noteworthy that among these 32 Hubble parameter measurements, 15 exhibit correlations, and we judiciously integrate these covariances into our analytical framework \footnote{The covariances can be accessed at the following link: \url{https://gitlab.com/mmoresco/CCcovariance/} \citep{Moresco:2020fbm,moresco2012improved,Moresco:2015cya,Moresco:2016mzx}.}, enhancing the precision of our analysis. The mean values and standard deviations of the observed Hubble parameter can be found in Table~\ref{table:H_CC_data}. It's noteworthy that for the last 15 redshift points marked with asterisks, non-zero covariances are present.

\begin{table}
\begin{center}
\begin{tabular}{|c|c|c|c|}
\hline
Sr. No. &$z$ & $H \pm \Delta H$ [km s$^{-1}$ Mpc$^{-1}$] & Refs. \\
\hline
1 & 0.07 & 69.0 $\pm$ 19.6 & \citep{zhang2014four} \\
2 & 0.09 & 69.0 $\pm$ 12.0 & \citep{Simon:2004tf} \\
3 & 0.12 & 68.6 $\pm$ 26.2 & \citep{zhang2014four} \\
4 & 0.17 & 83.0 $\pm$ 8.0 & \citep{Simon:2004tf} \\
5 & 0.2 & 72.9 $\pm$ 29.6 & \citep{zhang2014four} \\
6 & 0.27 & 77.0 $\pm$ 14.0 & \citep{Simon:2004tf} \\
7 & 0.28 & 88.8 $\pm$ 36.6 & \citep{zhang2014four} \\
8 & 0.4 & 95.0 $\pm$ 17.0 & \citep{Simon:2004tf} \\
9 & 0.47 & 89.0 $\pm$ 50.0 & \citep{Ratsimbazafy:2017vga} \\
10 & 0.48 & 97.0 $\pm$ 62.0 & \citep{stern2010cosmic} \\
11 & 0.75 & 98.8 $\pm$ 33.6 & \citep{Borghi:2021rft} \\
12 & 0.88 & 90.0 $\pm$ 40.0 & \citep{stern2010cosmic} \\
13 & 0.9 & 117.0 $\pm$ 23.0 & \citep{Simon:2004tf} \\
14 & 1.3 & 168.0 $\pm$ 17.0 & \citep{Simon:2004tf} \\
15 & 1.43 & 177.0 $\pm$ 18.0 & \citep{Simon:2004tf} \\
16 & 1.53 & 140.0 $\pm$ 14.0 & \citep{Simon:2004tf} \\
17 & 1.75 & 202.0 $\pm$ 40.0 & \citep{Simon:2004tf} \\
18 & 0.1791* & 74.91 $\pm$ 5.57 & \citep{Moresco:2020fbm} \\
19 & 0.1993* & 74.96 $\pm$ 6.37 & \citep{Moresco:2020fbm} \\
20 & 0.3519* & 82.78 $\pm$ 14.65 & \citep{Moresco:2020fbm} \\
21 & 0.3802* & 83.0 $\pm$ 14.3 & \citep{Moresco:2020fbm} \\
22 & 0.4004* & 76.97 $\pm$ 11.12 & \citep{Moresco:2020fbm} \\
23 & 0.4247* & 87.08 $\pm$ 12.47 & \citep{Moresco:2020fbm} \\
24 & 0.4497* & 92.78 $\pm$ 14.07 & \citep{Moresco:2020fbm} \\
25 & 0.4783* & 80.91 $\pm$ 10.23 & \citep{Moresco:2020fbm} \\
26 & 0.5929* & 103.8 $\pm$ 14.0 & \citep{Moresco:2020fbm} \\
27 & 0.6797* & 91.6 $\pm$ 9.5 & \citep{Moresco:2020fbm} \\
28 & 0.7812* & 104.5 $\pm$ 13.3 & \citep{Moresco:2020fbm} \\
29 & 0.8754* & 125.1 $\pm$ 17.1 & \citep{Moresco:2020fbm} \\
30 & 1.037* & 153.7 $\pm$ 20.4 & \citep{Moresco:2020fbm} \\
31 & 1.363* & 160.0 $\pm$ 32.9 & \citep{Moresco:2020fbm} \\
32 & 1.965* & 186.5 $\pm$ 49.8 & \citep{Moresco:2020fbm} \\
\hline
\end{tabular}
\end{center}
\caption{
The Hubble parameter's observed values, along with their 1$\sigma$ uncertainties, are documented at 32 redshift points, specifically from cosmic chronometers (CC) observations. Notably, at 15 of these redshift points marked with asterisks, correlations exist among each pair of observations, indicating the presence of covariances.
}
\label{table:H_CC_data}
\end{table}

\begin{figure}
\centering
\includegraphics[width=0.45\textwidth]{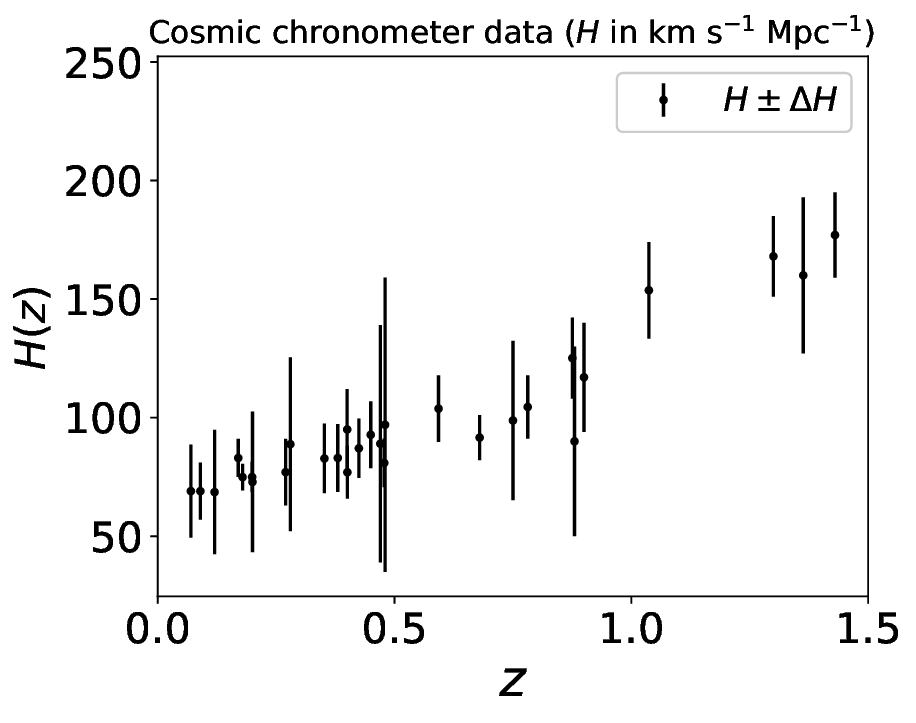}
\caption{
\label{fig:cosmic_chronometer_data}
Observed Hubble parameter data from cosmic chronometer observations with associated error bars representing 1$\sigma$ uncertainties.
}
\end{figure}

We have plotted the observed Hubble parameter data, along with their associated errors represented by error bars, in Figure~\ref{fig:cosmic_chronometer_data}.

Additionally, our study encompasses growth rate observations, as elucidated by \citep{Avila:2022xad}. These observations encapsulate a collection of 11 uncorrelated data points about the growth rate parameter $f$ within the redshift range $0.013 \leq z \leq 1.4$. The table provided in Table~\ref{table:f_data} presents the mean values and standard deviations of the logarithmic growth rate $f$. In the corresponding table, the survey names and cosmological tracer labels are provided alongside the observed $f$ values at various redshifts. It is assumed that within these data, there exists no correlation between different redshift points, or any existing correlations are deemed negligible.

\begin{table*}
\begin{center}
\begin{tabular}{|c|c|c|c|c|c|}
\hline
Sr. No. & Survey Name & $z$ & $f \pm \Delta f$ & References & Cosmological tracers \\
\hline
1 & ALFALFA & 0.013 & $0.56 \pm 0.07$ & \citep{Avila:2021dqv} & HI extragalactic sources \\
2 & 2dFGRS & 0.15 & $0.49 \pm 0.14$ & \citep{Hawkins:2002sg,Guzzo:2008ac} & galaxies \\
3 & GAMA & 0.18 & $0.49 \pm 0.12$ & \citep{Blake:2013nif} & multi-tracer: blue \& red galaxies \\
4 & WiggleZ   & 0.22 & $0.60 \pm 0.10$ & \citep{Blake_2011} & galaxies \\
5 & SDSS    & 0.35 & $0.70 \pm 0.18$ & \citep{SDSS:2006lmn} & luminous red galaxies (LRG) \\
6 & GAMA   & 0.38 & $0.66 \pm 0.09$ & \citep{Blake:2013nif} & multi-tracer: blue \& red galaxies \\
7 & WiggleZ & 0.41 & $0.70 \pm 0.07$ & \citep{Blake_2011} & galaxies \\
8 & 2SLAQ & 0.55 & $0.75 \pm 0.18$ & \citep{Ross:2006me} & LRG \& quasars \\
9 & WiggleZ & 0.60 & $0.73 \pm 0.07$ & \citep{Blake_2011} & galaxies \\
10 & VIMOS-VLT Deep Survey & 0.77 & $0.91 \pm 0.36$ & \citep{Guzzo:2008ac} & faint galaxies  \\
11 & 2QZ \& 2SLAQ & 1.40 & $0.90 \pm 0.24$ & \citep{daAngela:2006mf} & quasars \\
\hline
\end{tabular}
\end{center}
\caption{
The observed values of $f$ along with their 1$\sigma$ uncertainties are documented at 11 redshift points, aligning with various survey data.
}
\label{table:f_data}
\end{table*}

\begin{figure}
\centering
\includegraphics[width=0.45\textwidth]{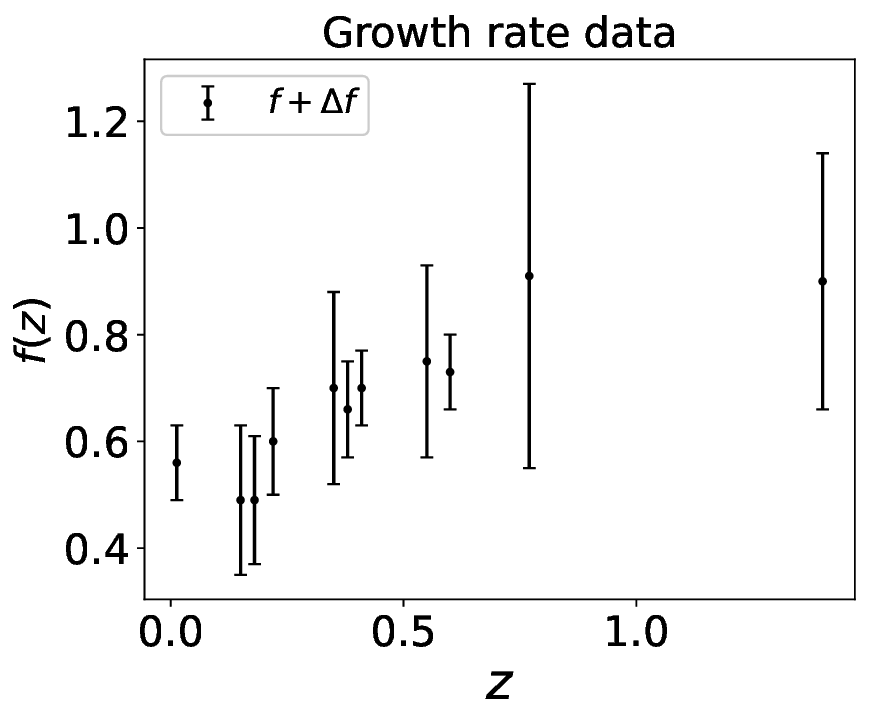}
\caption{
\label{fig:log_growth_f_data}
Logarithmic growth rate data with associated error bars representing 1$\sigma$ uncertainties.
}
\end{figure}

We have plotted the logarithmic growth rate data, along with their associated errors represented by error bars, in Figure~\ref{fig:log_growth_f_data}.

\section{Methodology}
\label{sec-method}

We employ analytical (posterior approach) Gaussian process regression (GPR) analysis \citep{williams1995gaussian,GpRasWil,Seikel_2012,PhysRevD.85.123530,Hwang:2022hla,Keeley:2020aym,Dinda:2022vmb,Dinda:2022jih,Perenon:2022fgw,Perenon:2021uom,OColgain:2021pyh,Banerjee:2023evd,Banerjee:2023rvg,Mukherjee:2022ujw,Mukherjee:2020vkx,Mukherjee:2020ytg,Li:2019nux,Ruiz-Zapatero:2022zpx,Ruiz-Zapatero:2022xbv} to reconstruct $H$ and $H'$ at specific redshift points from the observed Hubble parameter data. Similarly, utilizing analytical GPR, we reconstruct $f'$ at the same redshift points from the observed growth rate ($f$) data. We utilize analytical GPR for its convenience in handling observational data linearly related to GPR. Initially, we determine the posterior distributions of $H$, $f$, and their derivatives. Subsequently, we compute the mean and standard deviations at specific target redshift points, along with the covariances between each pair of these points. These computations are vital for determining $W_{\rm m0}$ i.e. $\Omega_{\rm m0} H_0^2$ through Eq.~\eqref{eq:Omega_m0_from_H_and_f}, where we require $H$, $f$, and their derivatives at the same redshift points. We align the observed redshift points of the $f$ data with our target redshift points of interest, except for the lowest redshift point ($z=0.013$). Particularly, we select 10 redshift points within the range $0.15 \leq z \leq 1.4$ based on the $f$ data, excluding the lowest one. These target redshift points are detailed in Table~\ref{table:z_target}.

\begin{table}
\begin{center}
\begin{tabular}{|c|c|}
\hline
Sr. No. & Redshift point \\
\hline
1 & 0.15 \\
2 & 0.18 \\
3 & 0.22 \\
4 & 0.35 \\
5 & 0.38 \\
6 & 0.41 \\
7 & 0.55 \\
8 & 0.60 \\
9 & 0.77 \\
10 & 1.40 \\
\hline
\end{tabular}
\end{center}
\caption{
Target redshift points.
}
\label{table:z_target}
\end{table}

This selection is motivated by several considerations. Firstly, to calculate $W_{\rm m0}$ using Eq.~\eqref{eq:Omega_m0_from_H_and_f}, we require all four quantities—$H$, $H'$, $f$, and $f'$—to be available at the same redshift points. Secondly, the chosen redshift range is situated within both the redshift ranges of the $H$ and $f$ data and employing interpolation techniques such as GPR can yield reliable results when interpolating from a specified range of redshifts to a subset range of redshifts. Thirdly, we have the flexibility to select either the observed $H$ data or the observed $f$ data to directly substitute into Equation~\eqref{eq:Omega_m0_from_H_and_f} for computing $\Omega_{\rm m0} H_0^2$. This approach allows us to incorporate at least one quantity directly obtained from observational data instead of relying solely on reconstructed values. For this purpose, we opt to use the observed $f$ data in Eq.~\eqref{eq:Omega_m0_from_H_and_f}.

As mentioned earlier, in our investigation, we opt for the posterior approach of Gaussian process regression (GPR) analysis for its efficiency in computational time and straightforward applicability. Let's delve briefly into the workings of GPR analysis and its application in reconstructing a function and its derivatives, specifically the first order, at target points along with associated errors from a given dataset.

Consider a dataset featuring $n$ observational data points denoted by vectors $X$ and $Y$, representing observation coordinates and mean values of a quantity, respectively. The dataset also incorporates observational errors through the covariance matrix $C$, denoted as $C = \text{Cov}[Y,Y]$. GPR analysis facilitates the prediction of mean values and covariances for the same quantity at different target points $X_*$, represented by vectors $Y_*$ and $\text{Cov}[Y_*,Y_*]$, leveraging a kernel covariance function and a mean function. In this context, we assume a zero mean function to eliminate model dependence, enhancing the versatility of our approach. The predicted values are computed through the expressions given by \citep{williams1995gaussian, GpRasWil, Seikel_2012, PhysRevD.85.123530}:

\begin{align}
\label{eq:GPR_mean_prediction}
Y_* &= K(X_*,X) \left[ K(X,X)+C \right]^{-1} Y, \\
\label{eq:GPR_cov_prediction}
\text{Cov}[Y_*,Y_*] &= K(X_*,X_*) \nonumber\\
& - K(X_*,X) \left[ K(X,X)+C \right]^{-1} K(X,X_*),
\end{align}

\noindent
where $K$ is the kernel matrix based on a specific kernel covariance function. We adopt the squared-exponential kernel, where the covariance between two arbitrary points $x_i$ and $x_j$ is expressed as:

\begin{equation}
k(x_i,x_j) = \sigma_f^2 e^{ -\frac{ (x_i-x_j)^2 }{2 l^2} },
\label{eq:kernel_SE}
\end{equation}

\noindent
where $\sigma_f$ and $l$ are the corresponding kernel hyperparameters, and we incorporate optimal values for these hyperparameters in predictions for Eqs. \eqref{eq:GPR_mean_prediction} and \eqref{eq:GPR_cov_prediction}. Determining these optimal values involves minimizing the negative log marginal likelihood ($\log P(Y|X)$), as presented in \citep{Seikel_2012}:

\begin{eqnarray}
\log P(Y|X) &=& -\frac{1}{2} Y^T \left[ K(X,X)+C \right]^{-1} Y \nonumber\\
&& -\frac{1}{2} \log |K(X,X)+C| -\frac{n}{2} \log{(2 \pi)},
\label{eq:log_marginal_likelihood}
\end{eqnarray}

\noindent
where $|K(X,X)+C|$ represents the determinant of the $K(X,X)+C$ matrix.

\begin{figure}
\centering
\includegraphics[width=0.45\textwidth]{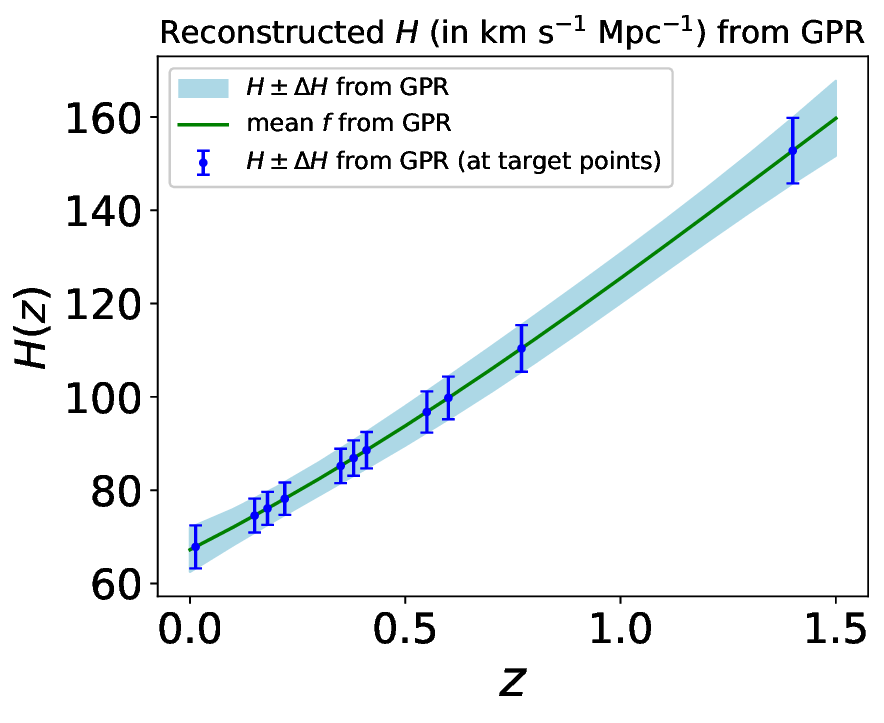}
\caption{
\label{fig:GPR_H_dH}
Reconstructed values of the Hubble parameter along with associated errors obtained from analytical Gaussian Process Regression (GPR). The green line and light-blue regions denote the mean function and 1$\sigma$ uncertainty region, respectively. Blue error bars represent the mean and standard deviation values of the reconstructed Hubble parameter at target redshift points mentioned in Table~\ref{table:z_target}.
}
\end{figure}

In Figure~\ref{fig:GPR_H_dH}, we have plotted the reconstructed values of the Hubble parameter and the associated errors obtained from analytical GPR. The green line and the light-blue regions represent the corresponding mean function and the 1$\sigma$ uncertainty region, respectively. The blue error bars represent the mean and standard deviation values of the reconstructed Hubble parameter at the target redshift points mentioned in Table~\ref{table:z_target}.

\begin{figure}
\centering
\includegraphics[width=0.45\textwidth]{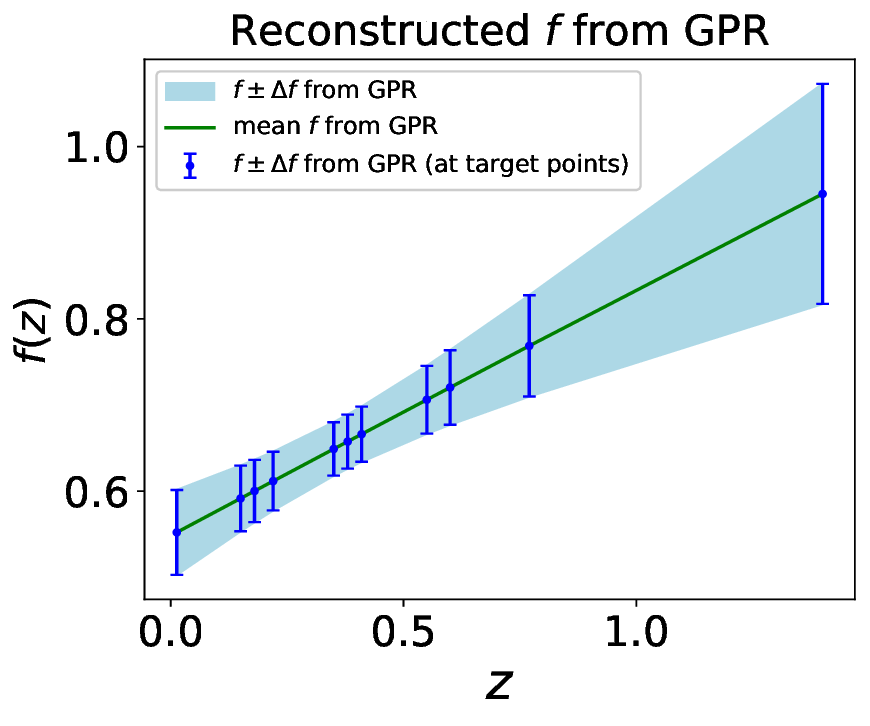}
\caption{
\label{fig:GPR_f_df}
The reconstructed values of the logarithmic growth rate ($f$) obtained from analytical Gaussian Process Regression (GPR) analysis. The green line represents the mean function, while the light-blue regions indicate the 1$\sigma$ uncertainty region. Blue error bars depict the mean and standard deviation values of the reconstructed $f$ at the specified target redshift points mentioned in Table~\ref{table:z_target}.
}
\end{figure}

In Figure~\ref{fig:GPR_f_df}, we have plotted the reconstructed values of $f$ and the associated errors obtained from the analytical GPR analysis. The green line and the light-blue regions represent the corresponding mean function and the 1$\sigma$ uncertainty region, respectively. The blue error bars represent the mean and standard deviation values of the reconstructed $f$ at the target redshift points mentioned in Table~\ref{table:z_target}.

Furthermore, GPR extends its predictive capabilities to the gradient of a quantity. The mean vector and covariance matrix corresponding to the first derivative are articulated by \citep{Seikel_2012}:

\begin{align}
\label{eq:derivative_mean_predictions}
Y'_* &= [K'(X,X_*)]^T \left[ K(X,X)+C \right]^{-1} Y, \\
\label{eq:derivative_cov_predictions}
\text{Cov}[Y'_*,Y'_*] &= K''(X_*,X_*) \nonumber\\
& - [K'(X,X_*)]^T \left[ K(X,X)+C \right]^{-1} K'(X,X_*),
\end{align}

\noindent
where prime and double prime denote the first and second derivatives, respectively. $k'(x,x_*)$ and $k''(x_*,x_*)$ represent the partial derivatives of the kernel function:

\begin{eqnarray}
k'(x,x_*) &=& \dfrac{\partial k(x,x_*)}{\partial x_*}, \nonumber\\
k''(x_*,x_*) &=& \dfrac{\partial ^2 k(x_*,x_*)}{\partial x_* \partial x_*},
\label{eq:kernel_derivatives}
\end{eqnarray}

\noindent
where the notation $k$ denotes the matrix element of the main matrix $K$. Additionally, the covariance matrix between the quantity and its first derivative is given by \citep{Seikel_2012}:

\begin{align}
\label{eq:cov_Y_Yp}
\text{Cov}[Y_*,Y'_*] &= K'(X_*,X_*) \nonumber\\
& - [K(X,X_*)]^T \left[ K(X,X)+C \right]^{-1} K'(X,X_*).
\end{align}

\begin{figure}
\centering
\includegraphics[width=0.45\textwidth]{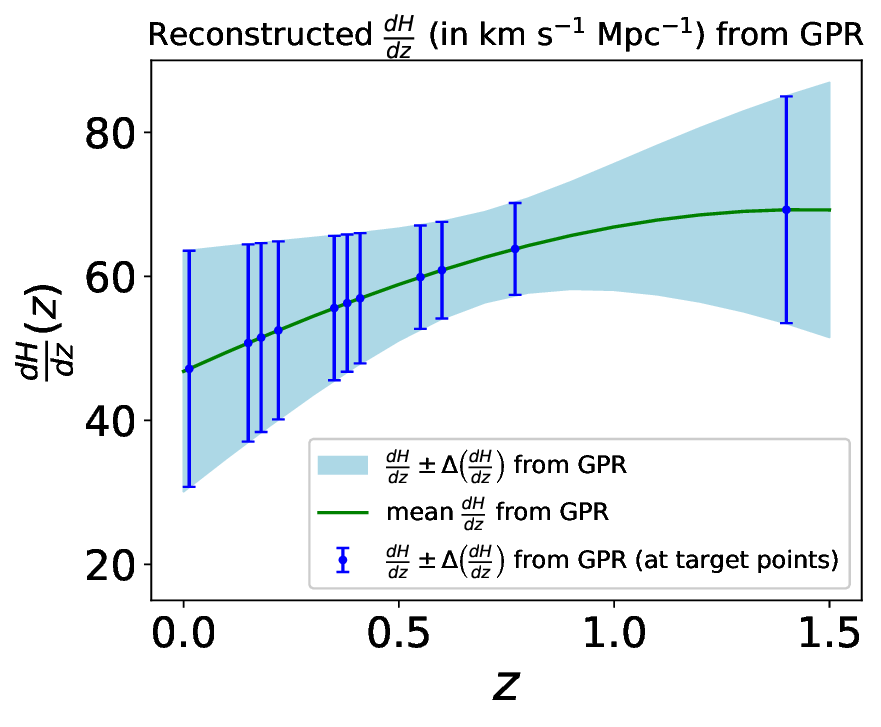}
\caption{
\label{fig:GPR_dHdz}
Reconstructed values of $H'$ i.e. $\frac{dH}{dz}$ obtained from analytical Gaussian Process Regression (GPR) analysis. The green line denotes the mean function, while the light-blue regions represent the 1$\sigma$ uncertainty region. Blue error bars indicate the mean and standard deviation values of the reconstructed $H'$ at the specified redshift points mentioned in Table~\ref{table:z_target}.
}
\end{figure}

In Figure~\ref{fig:GPR_dHdz}, we have plotted the reconstructed values of $H'$ i.e. $\frac{dH}{dz}$ and the associated errors obtained from analytical GPR. The green line and the light-blue regions represent the corresponding mean function and the 1$\sigma$ uncertainty region, respectively. The blue error bars represent the mean and standard deviation values of the reconstructed $H'$ at the target redshift points mentioned in Table~\ref{table:z_target}.

\begin{figure}
\centering
\includegraphics[width=0.45\textwidth]{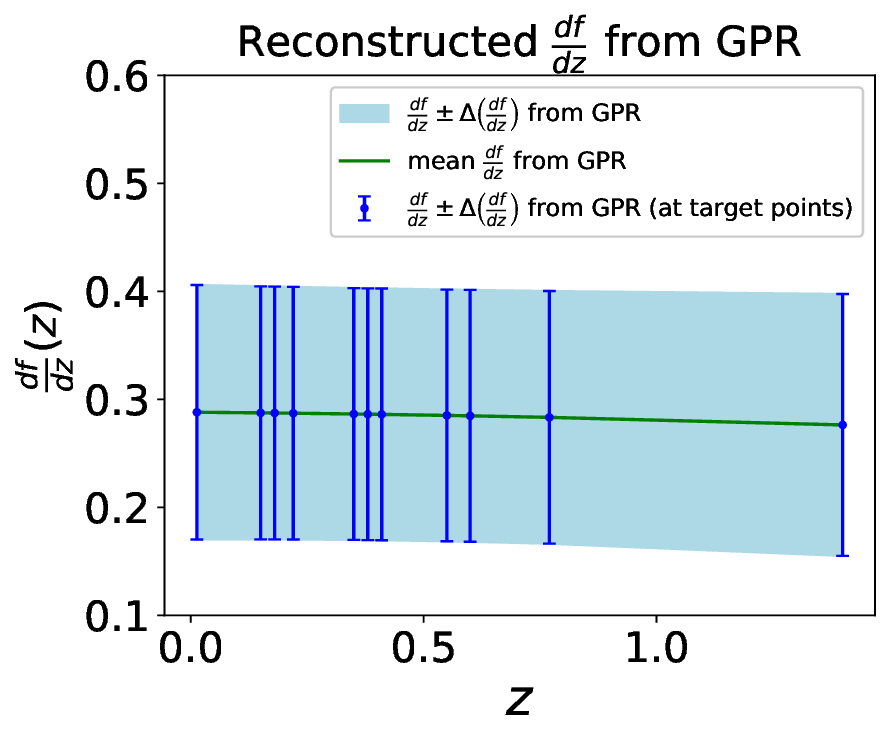}
\caption{
\label{fig:GPR_dfdz}
Reconstructed values of $f'$, denoted as $\frac{df}{dz}$, with associated errors. The green line represents the mean function, while the light-blue regions indicate the 1$\sigma$ uncertainty region obtained through analytical Gaussian Process Regression (GPR). Blue error bars denote the mean and standard deviation values of $f'$ at specified redshift points mentioned in Table~\ref{table:z_target}.
}
\end{figure}

In Figure \ref{fig:GPR_dfdz}, we present the reconstructed values of $f'$ i.e. $\frac{df}{dz}$ along with their associated errors obtained through analytical Gaussian Process Regression (GPR). The green line depicts the mean function, while the light-blue regions indicate the 1$\sigma$ uncertainty region. Blue error bars denote the mean and standard deviation values of the reconstructed $f'$ at the specified redshift points mentioned in Table~\ref{table:z_target}.

In summary, GPR proves to be a versatile tool, not only predicting the function and its derivatives but also providing insights into their covariations. This comprehensive predictive capability enhances the utility of GPR in our study, allowing for robust analyses and accurate reconstructions.

\section{Results}
\label{sec-result}

After conducting GPR analysis, we now possess the reconstructed values of $H$, $H'$, Var$[H]$, Var$[H']$, and Cov$[H,H']$ at each redshift point of interest, as mentioned earlier. Simultaneously, we have the reconstructed values of $f$, $f'$, Var$[f]$, Var$[f']$, and Cov$[f,f']$. Note that, we utilize the observed values of $f$ and Var$[f]$ in Eqs.~\eqref{eq:Omega_m0_from_H_and_f} and~\eqref{eq:var_Wm0} instead of their reconstructed counterparts from GPR. Although this step is not strictly necessary, it offers the advantage of incorporating true observational data directly, thereby mitigating potential errors that may arise during the reconstruction process, as mentioned earlier. However, a complication arises as follows: we need Cov$[f,f']$ in Eq.~\eqref{eq:var_Wm0} and we know these values between the reconstructed $f$ and $f'$ from GPR using Eq.~\eqref{eq:cov_Y_Yp}, but we do not know the values if we use observed $f$ instead of the reconstructed $f$. That means we do not know the covariances between the observed $f$ and the reconstructed $f'$.

To address this issue, we assume that the normalized covariance between the observed $f$ and the reconstructed $f'$ is equivalent to that between the reconstructed $f$ and the reconstructed $f'$. Here, the normalized covariance, denoted by $\rho [f,f']$, is defined as Cov$[f,f'] = \rho [f,f'] \cdot \Delta f \cdot \Delta f'$, where $\Delta A$ represents the standard deviation of a quantity $A$, given by $\Delta A = \sqrt{\text{Var}[A]}$. We use the errors from the observed $f$ data for $\Delta f$ and the reconstructed values from GPR for $\Delta f'$. This approach enables us to obtain covariances between the observed $f$ and the reconstructed $f'$.

The reconstruction of $H$ and $H'$, as well as the reconstruction of $f$ and $f'$, involves distinct datasets, leading to the absence of covariances between these two groups. Therefore, the only non-zero covariances are Cov$[H,H']$ and Cov$[f,f']$. This explains why we have exclusively taken these two covariances into account in Eq.~\eqref{eq:var_Wm0}. However, it is important to note that this assumption may not hold strictly true because theoretically, $H$ and $f$ are related through Eq.~\eqref{eq:diff_eqn_f}, implying correlations between $H$, $f$, and their derivatives. However, in our simplified methodology utilizing posterior GPR analysis, where the observed data is linearly related to GPR, imposing such correlations would require multi-tasking of GPR, which is challenging. Therefore, for the sake of simplicity, we assume either no correlation or negligible correlation \citep{Ruiz-Zapatero:2022xbv}.

Finally, utilizing these computed values, we derive $W_{\rm m0}(z_i)$ and $\Delta W_{\rm m0}(z_i)$ at each redshift point $z_i$ through Eqs.\eqref{eq:Omega_m0_from_H_and_f} and\eqref{eq:var_Wm0} respectively. From these results, we further obtain $\Omega_{\rm m0}h^2 (z_i)$ and $\Delta (\Omega_{\rm m0}h^2) (z_i)$ at each redshift by applying the relations:

\begin{eqnarray}
\Omega_{\rm m0}h^2 (z_i) &=& \frac{W_{\rm m0} (z_i)}{10^4 \left( \text{km s}^{-1} \text{Mpc}^{-1} \right)^2},
\label{eq:Omegam0_h2} \\
\Delta (\Omega_{\rm m0}h^2) (z_i) &=& \frac{\Delta W_{\rm m0} (z_i)}{10^4 \left( \text{km s}^{-1} \text{Mpc}^{-1} \right)^2},
\label{eq:delta_Om0_h2}
\end{eqnarray}

\noindent
where we express the present value of the Hubble parameter as $H_0=100~h$ km s$^{-1}$ Mpc$^{-1}$.

\begin{figure}
\centering
\includegraphics[width=0.45\textwidth]{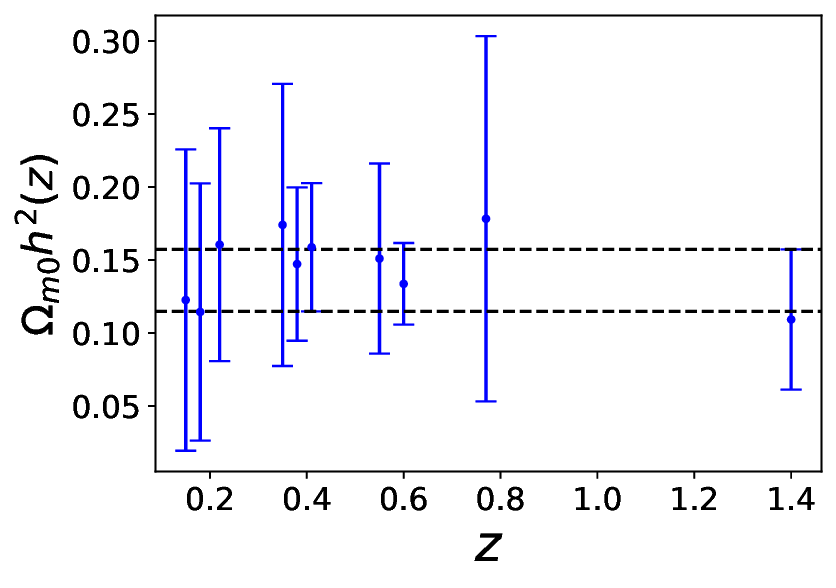}
\caption{
\label{fig:Omegam0h2}
The reconstruction process yields values for $\Omega_{\rm m0}h^2 (z_i)$ and $\Delta (\Omega_{\rm m0}h^2) (z_i)$ at 10 distinct redshift points denoted by $z_i$, depicted as blue error bars. The two horizontal dashed-black lines delineate the maximum allowable region for $\Omega_{\rm m0}h^2$ where all 1$\sigma$ error bars converge. The lower and upper horizontal black lines correspond to $\Omega_{\rm m0}h^2 \approx 0.115$ and $\Omega_{\rm m0}h^2 \approx 0.157$, respectively.
}
\end{figure}

In Figure~\ref{fig:Omegam0h2}, the reconstructed values of $\Omega_{\rm m0}h^2 (z_i)$ and $\Delta (\Omega_{\rm m0}h^2) (z_i)$ are depicted with blue error bars. The two horizontal dashed-black lines define the maximum region that all error bars commonly share. This region is crucial, as any single value of $\Omega_{\rm m0}h^2$ falling within it is supported by all 10 reconstructed values of $\Omega_{\rm m0}h^2 (z_i)$ at a 1$\sigma$ confidence level. It's worth noting that the lower and upper horizontal black lines correspond to $\Omega_{\rm m0}h^2 \approx 0.115$ and $\Omega_{\rm m0}h^2 \approx 0.157$ respectively. At a 2$\sigma$ confidence level, this region would expand, and so forth.

According to the first-order Newtonian perturbations within the background FLRW metric, $\Omega_{\rm m0}h^2$ should be a constant. Therefore, if observations align completely with this underlying theory and the first-order Newtonian perturbation theory, the derived values of $\Omega_{\rm m0}h^2$ should be a constant regardless of the redshift point of the observed data. However, considering the presence of error bars, the data provides a region of possible values for $\Omega_{\rm m0}h^2$ at each observed redshift, depending on the confidence interval. At a specific confidence level, all observed regions of $\Omega_{\rm m0}h^2$ at different redshift points should intersect. The presence of an overlap region between the horizontal dashed-black lines in Figure~\ref{fig:Omegam0h2} confirms the consistency between the first-order Newtonian perturbations, the background FLRW metric, and the cosmic chronometers and growth rate data. This also underscores the utility of Eq.~\eqref{eq:Omega_m0_from_H_and_f} for a simultaneous consistency test of the FLRW background metric and first-order Newtonian perturbations.

Given that the reconstructed values of $\Omega_{\rm m0}h^2 (z_i)$ at each redshift $z_i$ align with the notion that $\Omega_{\rm m0}h^2$ remains constant across these points, we can view these reconstructions as 10 distinct measurements of the same quantity, $\Omega_{\rm m0}h^2$, each with associated errors. With this interpretation, we employ standard parameter estimation techniques to calculate the $\Omega_{\rm m0}h^2$ parameter and its associated error. The mean value and variance of the $\Omega_{\rm m0}h^2$ parameter are computed using the following relations \citep{Cao:2019kgn,Liu:2020pfa}

\begin{eqnarray}
\Omega_{\rm m0}h^2 &=& \frac{ \sum_{i} \frac{\Omega_{\rm m0}h^2 (z_i)}{\text{Var}[\Omega_{\rm m0}h^2](z_i)} }{ \sum_{j} \frac{1}{\text{Var}[\Omega_{\rm m0}h^2](z_j)} },
\label{eq:Omegam0_h2_mean} \\
\text{Var}[\Omega_{\rm m0}h^2] &=& \frac{ 1 }{ \sum_{i} \frac{1}{\text{Var}[\Omega_{\rm m0}h^2](z_i)} },
\label{eq:Variance_Om0_h2}
\end{eqnarray}

\noindent
respectively and the standard deviation is determined as $\Delta (\Omega_{\rm m0}h^2)=\sqrt{\text{Var}[\Omega_{\rm m0}h^2]}$. These equations emphasize assigning higher weight to $\Omega_{\rm m0}h^2 (z_i)$ with lower variance. The estimated value of $\Omega_{\rm m0}h^2$ and its associated 1$\sigma$ error are presented in Table~\ref{table:Om0h2_mean_std}.

\begin{table}
\begin{center}
\begin{tabular}{ |c| }
\hline
$\Omega_{\rm m0}h^2 = 0.139 \pm 0.017$ \\
\hline
\end{tabular}
\end{center}
\caption{
The estimated value of the $\Omega_{\rm m0}h^2$ parameter and the associated 1$\sigma$ error.
}
\label{table:Om0h2_mean_std}
\end{table}

Now to estimate the value of $\Omega_{\rm m0}$, we have to break the degeneracy in $\Omega_{\rm m0}h^2$ by using the measured value of $H_0$. For the $H_0$ value, we first compute it from the cosmic chronometer data itself. What we do we use the same procedure GPR analysis to compute $H$ at $z=0$. We find the value as $H_0=67.2\pm4.7$. We also consider two other values of $H_0$ from two different observations. One is from the tip of the Red Giant Branch (tRGB) measurements which correspond to $H_0 \approx 69.8\pm1.9$ \cite{Freedman:2019jwv}. The other one is from the SHOES measurement which corresponds to $H_0=73.2\pm1.3$ \cite{Riess:2020fzl}. From these $H_0$ values, we compute $h$ and $\Delta h$ using the relations given as

\begin{eqnarray}
h &=& \frac{ H_0 }{ 100 \text{ km s}^{-1} \text{ Mpc}^{-1} },
\label{eq:h_from_H0} \\
\Delta h &=& \frac{ \Delta H_0 }{ 100 \text{ km s}^{-1} \text{ Mpc}^{-1} },
\label{eq:std_h_from_std_H0}
\end{eqnarray}

\noindent
respectively. Now, from the values of $\Omega_{\rm m0}h^2$, $\Delta (\Omega_{\rm m0}h^2)$, $h$, and $\Delta h$, we compute $\Omega_{\rm m0}$, $\text{Var}[\Omega_{\rm m0}]$ using the relations

\begin{eqnarray}
\Omega_{\rm m0} &=& \frac{ (\Omega_{\rm m0}h^2) }{ h^2 },
\label{eq:final_Om0} \\
\text{Var}[\Omega_{\rm m0}] &=& \left[ \frac{\partial \Omega_{\rm m0}}{\partial (\Omega_{\rm m0}h^2)} \right]^2 \text{Var}[\Omega_{\rm m0}h^2] + \left[ \frac{\partial \Omega_{\rm m0}}{\partial h} \right]^2 \text{Var}[h] \nonumber\\
&& = \frac{1}{h^4} \text{Var}[\Omega_{\rm m0}h^2] + \frac{4(\Omega_{\rm m0}h^2)^2}{h^6} \text{Var}[h],
\label{eq:final_std_Om0}
\end{eqnarray}

\noindent
respectively and we find $\Delta \Omega_{\rm m0}$ as $\Delta \Omega_{\rm m0}=\sqrt{\text{Var}[\Omega_{\rm m0}]}$. We list all these estimated values in Table~\ref{table:H0_h_Om0}.

\begin{table}
\begin{center}
\begin{tabular}{ |c|c|c| }
\hline
Observation & $h\pm \Delta h$ & $\Omega_{\rm m0}\pm \Delta \Omega_{\rm m0}$ \\
\hline
CC (using GPR) & $0.672\pm 0.047$ & $0.308\pm 0.057$ \\
\hline
tRGB & $0.698\pm 0.019$ & $0.285\pm 0.038$ \\
\hline
SHOES & $0.732\pm 0.013$ & $0.259\pm 0.033$ \\
\hline
\end{tabular}
\end{center}
\caption{
The estimated values of the $\Omega_{\rm m0}$ parameter and the associated 1$\sigma$ error correspond to three different sets of $h$ and $\Delta h$.
}
\label{table:H0_h_Om0}
\end{table}

\begin{figure}
\centering
\includegraphics[width=0.45\textwidth]{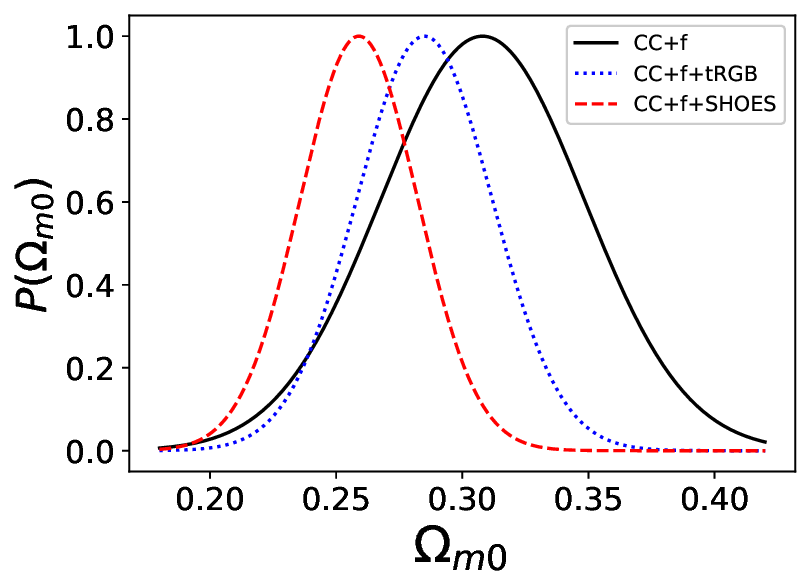}
\caption{
\label{fig:Omegam0_pdf}
Probabilty distribution for $\Omega_{\rm m0}$. The continuous black, dotted blue, and dashed red lines represent the constraints on $\Omega_{\rm m0}$ for the combinations of 'CC+f', 'CC+f+tRGB', and 'CC+f+SHOES' data sets, respectively.
}
\end{figure}

From the estimated values of the $\Omega_{\rm m0}$ and $\Delta \Omega_{\rm m0}$, we plot the Gaussian probabilty distribution for $\Omega_{\rm m0}$ (denoted as $P(\Omega_{\rm m0})$) in Figure~\ref{fig:Omegam0_pdf}. The solid-black, dotted-blue, and dashed-red lines correspond to the constraints on $\Omega_{\rm m0}$ for the combinations of 'CC+f', 'CC+f+tRGB', 'CC+f+SHOES' data sets respectively. We see that the higher the values of $H_0$, the lower the values of $\Omega_{\rm m0}$.

\section{Conclusion}
\label{sec-conclusion}

In this investigation, we amalgamate Hubble parameter data from cosmic chronometers (CC) observations with growth rate data obtained from growth rate (f) observations to derive constraints on the $\Omega_{\rm m0}$ parameter. Formulating a consistency relation for the combined parameter $\Omega_{\rm m0}h^2$ in terms of the Hubble parameter, its derivative, the growth rate ($f$), and the derivative of $f$ at a specific redshift $z$, we base our analysis on the assumption of a flat FLRW metric governing the background expansion of the Universe and the first-order Newtonian perturbation theory for the evolution of matter inhomogeneity. This relation serves as a valuable tool to scrutinize the consistency of the FLRW metric and first-order perturbations within the Newtonian perturbation theory, remaining independent of any particular cosmological model or parametrization.

Moreover, our investigation showcases the potential of analytical Gaussian Process (GP) cosmography as a valuable methodological advancement. By utilizing the differentiable properties of GPs, we derive the necessary quantities analytically, offering significant computational advantages over traditional numerical methods. The analytical approach not only simplifies the computational burden but also provides deeper interpretability of results, facilitating a more comprehensive understanding of cosmological phenomena.

Employing Gaussian process regression (GPR) analysis, we compute the Hubble parameter, its derivative, and associated errors from cosmic chronometer observations. Similarly, we employ GPR to determine $f$, $f'$, and their associated errors from growth rate measurements. Utilizing these reconstructed values in conjunction with the consistency relation, we constrain the $\Omega_{\rm m0}h^2$ parameter, resulting in $\Omega_{\rm m0}h^2=0.139\pm0.017$.

Subsequently, leveraging $H_0$ observations, we further constrain the $\Omega_{\rm m0}$ parameter. Initially, we use GPR to compute $H_0$ directly from cosmic chronometer observations, yielding $\Omega_{\rm m0}=0.308\pm0.057$ for $H_0=67.2\pm4.7$. We then consider the tip of the Red Giant Branch (tRGB) observation, corresponding to $H_0\approx69.8\pm1.9$, resulting in $\Omega_{\rm m0}=0.285\pm0.038$. Finally, utilizing the SHOES measurement of $H_0$ ($H_0=73.2\pm1.3$), we find $\Omega_{\rm m0}=0.259\pm0.033$. Notably, the study reveals an inverse correlation between the mean values of $H_0$ and $\Omega_{\rm m0}$.

In summary, this investigation confines the $\Omega_{\rm m0}$ parameter by integrating cosmic chronometers and growth rate observations, with or without additional Hubble constant measurements. Importantly, this is achieved in a completely cosmological model-independent manner, facilitated by the consistency relation governing the $\Omega_{\rm m0}h^2$ parameter. The model-independent methodology applied here, namely analytical Gaussian process cosmography, plays a crucial role in deriving constraints on $\Omega_{\rm m0}$ without reliance on specific cosmological models or parametrizations. By utilizing analytical Gaussian process regression to reconstruct cosmological observables, such as the Hubble parameter and growth rate, this approach ensures robust and interpretable results that align with the background FLRW metric and the first-order evolution of matter inhomogeneity in the Newtonian cosmological perturbation theory.

\section*{Data Availability Statement}
This manuscript has no associated data or the data will not be deposited. [Authors’ comment: This research uses publicly available cosmological data and wherever used cited properly.]

\begin{acknowledgements}
The author would like to acknowledge IISER Kolkata for its financial support through the postdoctoral fellowship. The revision for this work was carried out during the author's tenure at UWC, partially supported by the South African Radio Astronomy Observatory and  National Research Foundation (Grant No. 75415).
\end{acknowledgements}












\bibliographystyle{JHEP}
\bibliography{refsOmega}

\providecommand{\href}[2]{#2}\begingroup\raggedright\begin{thebibliography}{100}

\bibitem{SupernovaCosmologyProject:1997zqe}
{\bf Supernova Cosmology Project} Collaboration, S.~Perlmutter et~al., {\it
  {Discovery of a supernova explosion at half the age of the Universe and its
  cosmological implications}},  {\em Nature} {\bf 391} (1998) 51--54,
  [\href{http://arxiv.org/abs/astro-ph/9712212}{{\tt astro-ph/9712212}}].

\bibitem{SupernovaSearchTeam:1998fmf}
{\bf Supernova Search Team} Collaboration, A.~G. Riess et~al., {\it
  {Observational evidence from supernovae for an accelerating universe and a
  cosmological constant}},  {\em Astron. J.} {\bf 116} (1998) 1009--1038,
  [\href{http://arxiv.org/abs/astro-ph/9805201}{{\tt astro-ph/9805201}}].

\bibitem{SupernovaCosmologyProject:1998vns}
{\bf Supernova Cosmology Project} Collaboration, S.~Perlmutter et~al., {\it
  {Measurements of $\Omega$ and $\Lambda$ from 42 high redshift supernovae}},
  {\em Astrophys. J.} {\bf 517} (1999) 565--586,
  [\href{http://arxiv.org/abs/astro-ph/9812133}{{\tt astro-ph/9812133}}].

\bibitem{2011NatPh...7Q.833W}
A.~{Wright}, {\it {Nobel Prize 2011: Perlmutter, Schmidt \& Riess}},  {\em
  Nature Physics} {\bf 7} (Nov., 2011) 833.

\bibitem{Linden2009CosmologicalPE}
S.~Linden, J.~M. Virey, and A.~Tilquin, {\it Cosmological parameter extraction
  and biases from type ia supernova magnitude evolution},  {\em Astronomy and
  Astrophysics} {\bf 506} (2009) 1095--1105.

\bibitem{Camarena:2019rmj}
D.~Camarena and V.~Marra, {\it {A new method to build the (inverse) distance
  ladder}},  {\em Mon. Not. Roy. Astron. Soc.} {\bf 495} (2020), no.~3
  2630--2644, [\href{http://arxiv.org/abs/1910.14125}{{\tt arXiv:1910.14125}}].

\bibitem{Pan-STARRS1:2017jku}
{\bf Pan-STARRS1} Collaboration, D.~M. Scolnic et~al., {\it {The Complete
  Light-curve Sample of Spectroscopically Confirmed SNe Ia from Pan-STARRS1 and
  Cosmological Constraints from the Combined Pantheon Sample}},  {\em
  Astrophys. J.} {\bf 859} (2018), no.~2 101,
  [\href{http://arxiv.org/abs/1710.00845}{{\tt arXiv:1710.00845}}].

\bibitem{Camlibel:2020xbn}
A.~K. \c{C}aml\i{}bel, I.~Semiz, and M.~A. Feyizo\u{g}lu, {\it {Pantheon update
  on a model-independent analysis of cosmological supernova data}},  {\em
  Class. Quant. Grav.} {\bf 37} (2020), no.~23 235001,
  [\href{http://arxiv.org/abs/2001.04408}{{\tt arXiv:2001.04408}}].

\bibitem{Planck:2013pxb}
{\bf Planck} Collaboration, P.~A.~R. Ade et~al., {\it {Planck 2013 results.
  XVI. Cosmological parameters}},  {\em Astron. Astrophys.} {\bf 571} (2014)
  A16, [\href{http://arxiv.org/abs/1303.5076}{{\tt arXiv:1303.5076}}].

\bibitem{Planck:2015fie}
{\bf Planck} Collaboration, P.~A.~R. Ade et~al., {\it {Planck 2015 results.
  XIII. Cosmological parameters}},  {\em Astron. Astrophys.} {\bf 594} (2016)
  A13, [\href{http://arxiv.org/abs/1502.01589}{{\tt arXiv:1502.01589}}].

\bibitem{Planck:2018vyg}
{\bf Planck} Collaboration, N.~Aghanim et~al., {\it {Planck 2018 results. VI.
  Cosmological parameters}},  {\em Astron. Astrophys.} {\bf 641} (2020) A6,
  [\href{http://arxiv.org/abs/1807.06209}{{\tt arXiv:1807.06209}}]. [Erratum:
  Astron.Astrophys. 652, C4 (2021)].

\bibitem{Jimenez:2001gg}
R.~Jimenez and A.~Loeb, {\it {Constraining cosmological parameters based on
  relative galaxy ages}},  {\em Astrophys. J.} {\bf 573} (2002) 37--42,
  [\href{http://arxiv.org/abs/astro-ph/0106145}{{\tt astro-ph/0106145}}].

\bibitem{Pinho:2018unz}
A.~M. Pinho, S.~Casas, and L.~Amendola, {\it {Model-independent reconstruction
  of the linear anisotropic stress $\eta$}},  {\em JCAP} {\bf 11} (2018) 027,
  [\href{http://arxiv.org/abs/1805.00027}{{\tt arXiv:1805.00027}}].

\bibitem{Cao:2023eja}
S.~Cao and B.~Ratra, {\it {H0=69.8\ensuremath{\pm}1.3\,\,km\,s-1\,Mpc-1,
  \ensuremath{\Omega}m0=0.288\ensuremath{\pm}0.017, and other constraints from
  lower-redshift, non-CMB, expansion-rate data}},  {\em Phys. Rev. D} {\bf 107}
  (2023), no.~10 103521, [\href{http://arxiv.org/abs/2302.14203}{{\tt
  arXiv:2302.14203}}].

\bibitem{BOSS:2016wmc}
{\bf BOSS} Collaboration, S.~Alam et~al., {\it {The clustering of galaxies in
  the completed SDSS-III Baryon Oscillation Spectroscopic Survey: cosmological
  analysis of the DR12 galaxy sample}},  {\em Mon. Not. Roy. Astron. Soc.} {\bf
  470} (2017), no.~3 2617--2652, [\href{http://arxiv.org/abs/1607.03155}{{\tt
  arXiv:1607.03155}}].

\bibitem{eBOSS:2020yzd}
{\bf eBOSS} Collaboration, S.~Alam et~al., {\it {Completed SDSS-IV extended
  Baryon Oscillation Spectroscopic Survey: Cosmological implications from two
  decades of spectroscopic surveys at the Apache Point Observatory}},  {\em
  Phys. Rev. D} {\bf 103} (2021), no.~8 083533,
  [\href{http://arxiv.org/abs/2007.08991}{{\tt arXiv:2007.08991}}].

\bibitem{Hou:2020rse}
J.~Hou et~al., {\it {The Completed SDSS-IV extended Baryon Oscillation
  Spectroscopic Survey: BAO and RSD measurements from anisotropic clustering
  analysis of the Quasar Sample in configuration space between redshift 0.8 and
  2.2}},  {\em Mon. Not. Roy. Astron. Soc.} {\bf 500} (2020), no.~1 1201--1221,
  [\href{http://arxiv.org/abs/2007.08998}{{\tt arXiv:2007.08998}}].

\bibitem{Peebles:2002gy}
P.~J.~E. Peebles and B.~Ratra, {\it {The Cosmological Constant and Dark
  Energy}},  {\em Rev. Mod. Phys.} {\bf 75} (2003) 559--606,
  [\href{http://arxiv.org/abs/astro-ph/0207347}{{\tt astro-ph/0207347}}].

\bibitem{Copeland:2006wr}
E.~J. Copeland, M.~Sami, and S.~Tsujikawa, {\it {Dynamics of dark energy}},
  {\em Int. J. Mod. Phys. D} {\bf 15} (2006) 1753--1936,
  [\href{http://arxiv.org/abs/hep-th/0603057}{{\tt hep-th/0603057}}].

\bibitem{Yoo:2012ug}
J.~Yoo and Y.~Watanabe, {\it {Theoretical Models of Dark Energy}},  {\em Int.
  J. Mod. Phys. D} {\bf 21} (2012) 1230002,
  [\href{http://arxiv.org/abs/1212.4726}{{\tt arXiv:1212.4726}}].

\bibitem{Lonappan:2017lzt}
A.~I. Lonappan, S.~Kumar, Ruchika, B.~R. Dinda, and A.~A. Sen, {\it {Bayesian
  evidences for dark energy models in light of current observational data}},
  {\em Phys. Rev. D} {\bf 97} (2018), no.~4 043524,
  [\href{http://arxiv.org/abs/1707.00603}{{\tt arXiv:1707.00603}}].

\bibitem{Dinda:2017swh}
B.~R. Dinda, {\it {Probing dark energy using convergence power spectrum and
  bi-spectrum}},  {\em JCAP} {\bf 09} (2017) 035,
  [\href{http://arxiv.org/abs/1705.00657}{{\tt arXiv:1705.00657}}].

\bibitem{Dinda:2018uwm}
B.~R. Dinda, A.~A. Sen, and T.~R. Choudhury, {\it {Dark energy constraints from
  the 21\textasciitilde{}cm intensity mapping surveys with SKA1}},
  \href{http://arxiv.org/abs/1804.11137}{{\tt arXiv:1804.11137}}.

\bibitem{Clifton:2011jh}
T.~Clifton, P.~G. Ferreira, A.~Padilla, and C.~Skordis, {\it {Modified Gravity
  and Cosmology}},  {\em Phys. Rept.} {\bf 513} (2012) 1--189,
  [\href{http://arxiv.org/abs/1106.2476}{{\tt arXiv:1106.2476}}].

\bibitem{Koyama:2015vza}
K.~Koyama, {\it {Cosmological Tests of Modified Gravity}},  {\em Rept. Prog.
  Phys.} {\bf 79} (2016), no.~4 046902,
  [\href{http://arxiv.org/abs/1504.04623}{{\tt arXiv:1504.04623}}].

\bibitem{Tsujikawa:2010zza}
S.~Tsujikawa, {\it {Modified gravity models of dark energy}},  {\em Lect. Notes
  Phys.} {\bf 800} (2010) 99--145, [\href{http://arxiv.org/abs/1101.0191}{{\tt
  arXiv:1101.0191}}].

\bibitem{Joyce:2016vqv}
A.~Joyce, L.~Lombriser, and F.~Schmidt, {\it {Dark Energy Versus Modified
  Gravity}},  {\em Ann. Rev. Nucl. Part. Sci.} {\bf 66} (2016) 95--122,
  [\href{http://arxiv.org/abs/1601.06133}{{\tt arXiv:1601.06133}}].

\bibitem{Dinda:2017lpz}
B.~R. Dinda, M.~Wali~Hossain, and A.~A. Sen, {\it {Observed galaxy power
  spectrum in cubic Galileon model}},  {\em JCAP} {\bf 01} (2018) 045,
  [\href{http://arxiv.org/abs/1706.00567}{{\tt arXiv:1706.00567}}].

\bibitem{Dinda:2018eyt}
B.~R. Dinda, {\it {Weak lensing probe of cubic Galileon model}},  {\em JCAP}
  {\bf 06} (2018) 017, [\href{http://arxiv.org/abs/1801.01741}{{\tt
  arXiv:1801.01741}}].

\bibitem{Zhang:2020qkd}
J.~Zhang, B.~R. Dinda, M.~W. Hossain, A.~A. Sen, and W.~Luo, {\it {Study of
  cubic Galileon gravity using $N$-body simulations}},  {\em Phys. Rev. D} {\bf
  102} (2020), no.~4 043510, [\href{http://arxiv.org/abs/2004.12659}{{\tt
  arXiv:2004.12659}}].

\bibitem{Dinda:2022ixi}
B.~R. Dinda, M.~W. Hossain, and A.~A. Sen, {\it {21 cm power spectrum in
  interacting cubic Galileon model}},
  \href{http://arxiv.org/abs/2208.11560}{{\tt arXiv:2208.11560}}.

\bibitem{Bassi:2023vaq}
A.~Bassi, B.~R. Dinda, and A.~A. Sen, {\it {21 cm Power Spectrum for Bimetric
  Gravity and its Detectability with SKA1-Mid Telescope}},
  \href{http://arxiv.org/abs/2306.03875}{{\tt arXiv:2306.03875}}.

\bibitem{Nojiri:2010wj}
S.~Nojiri and S.~D. Odintsov, {\it {Unified cosmic history in modified gravity:
  from F(R) theory to Lorentz non-invariant models}},  {\em Phys. Rept.} {\bf
  505} (2011) 59--144, [\href{http://arxiv.org/abs/1011.0544}{{\tt
  arXiv:1011.0544}}].

\bibitem{Nojiri:2017ncd}
S.~Nojiri, S.~D. Odintsov, and V.~K. Oikonomou, {\it {Modified Gravity Theories
  on a Nutshell: Inflation, Bounce and Late-time Evolution}},  {\em Phys.
  Rept.} {\bf 692} (2017) 1--104, [\href{http://arxiv.org/abs/1705.11098}{{\tt
  arXiv:1705.11098}}].

\bibitem{Bamba:2012cp}
K.~Bamba, S.~Capozziello, S.~Nojiri, and S.~D. Odintsov, {\it {Dark energy
  cosmology: the equivalent description via different theoretical models and
  cosmography tests}},  {\em Astrophys. Space Sci.} {\bf 342} (2012) 155--228,
  [\href{http://arxiv.org/abs/1205.3421}{{\tt arXiv:1205.3421}}].

\bibitem{Lee:2022cyh}
B.-H. Lee, W.~Lee, E.~O. Colg\'ain, M.~M. Sheikh-Jabbari, and S.~Thakur, {\it
  {Is local H $_{0}$ at odds with dark energy EFT?}},  {\em JCAP} {\bf 04}
  (2022), no.~04 004, [\href{http://arxiv.org/abs/2202.03906}{{\tt
  arXiv:2202.03906}}].

\bibitem{Carroll:2000fy}
S.~M. Carroll, {\it {The Cosmological constant}},  {\em Living Rev. Rel.} {\bf
  4} (2001) 1, [\href{http://arxiv.org/abs/astro-ph/0004075}{{\tt
  astro-ph/0004075}}].

\bibitem{Zlatev:1998tr}
I.~Zlatev, L.-M. Wang, and P.~J. Steinhardt, {\it {Quintessence, cosmic
  coincidence, and the cosmological constant}},  {\em Phys. Rev. Lett.} {\bf
  82} (1999) 896--899, [\href{http://arxiv.org/abs/astro-ph/9807002}{{\tt
  astro-ph/9807002}}].

\bibitem{Sahni:1999gb}
V.~Sahni and A.~A. Starobinsky, {\it {The Case for a positive cosmological
  Lambda term}},  {\em Int. J. Mod. Phys. D} {\bf 9} (2000) 373--444,
  [\href{http://arxiv.org/abs/astro-ph/9904398}{{\tt astro-ph/9904398}}].

\bibitem{Velten:2014nra}
H.~Velten, R.~vom Marttens, and W.~Zimdahl, {\it {Aspects of the cosmological
  \textquotedblleft{}coincidence problem\textquotedblright{}}},  {\em Eur.
  Phys. J. C} {\bf 74} (2014), no.~11 3160,
  [\href{http://arxiv.org/abs/1410.2509}{{\tt arXiv:1410.2509}}].

\bibitem{Malquarti:2003hn}
M.~Malquarti, E.~J. Copeland, and A.~R. Liddle, {\it {K-essence and the
  coincidence problem}},  {\em Phys. Rev. D} {\bf 68} (2003) 023512,
  [\href{http://arxiv.org/abs/astro-ph/0304277}{{\tt astro-ph/0304277}}].

\bibitem{DiValentino:2021izs}
E.~Di~Valentino, O.~Mena, S.~Pan, L.~Visinelli, W.~Yang, A.~Melchiorri, D.~F.
  Mota, A.~G. Riess, and J.~Silk, {\it {In the Realm of the Hubble tension $-$
  a Review of Solutions}},  \href{http://arxiv.org/abs/2103.01183}{{\tt
  arXiv:2103.01183}}.

\bibitem{Krishnan:2021dyb}
C.~Krishnan, R.~Mohayaee, E.~O. Colg\'ain, M.~M. Sheikh-Jabbari, and L.~Yin,
  {\it {Does Hubble tension signal a breakdown in FLRW cosmology?}},  {\em
  Class. Quant. Grav.} {\bf 38} (2021), no.~18 184001,
  [\href{http://arxiv.org/abs/2105.09790}{{\tt arXiv:2105.09790}}].

\bibitem{Vagnozzi:2019ezj}
S.~Vagnozzi, {\it {New physics in light of the $H_0$ tension: An alternative
  view}},  {\em Phys. Rev. D} {\bf 102} (2020), no.~2 023518,
  [\href{http://arxiv.org/abs/1907.07569}{{\tt arXiv:1907.07569}}].

\bibitem{Dinda:2021ffa}
B.~R. Dinda, {\it {Cosmic expansion parametrization: Implication for curvature
  and H0 tension}},  {\em Phys. Rev. D} {\bf 105} (2022), no.~6 063524,
  [\href{http://arxiv.org/abs/2106.02963}{{\tt arXiv:2106.02963}}].

\bibitem{DiValentino:2020vvd}
E.~Di~Valentino et~al., {\it {Cosmology Intertwined III: $f \sigma_8$ and
  $S_8$}},  {\em Astropart. Phys.} {\bf 131} (2021) 102604,
  [\href{http://arxiv.org/abs/2008.11285}{{\tt arXiv:2008.11285}}].

\bibitem{Abdalla:2022yfr}
E.~Abdalla et~al., {\it {Cosmology intertwined: A review of the particle
  physics, astrophysics, and cosmology associated with the cosmological
  tensions and anomalies}},  {\em JHEAp} {\bf 34} (2022) 49--211,
  [\href{http://arxiv.org/abs/2203.06142}{{\tt arXiv:2203.06142}}].

\bibitem{Douspis:2018xlj}
M.~Douspis, L.~Salvati, and N.~Aghanim, {\it {On the Tension between Large
  Scale Structures and Cosmic Microwave Background}},  {\em PoS} {\bf EDSU2018}
  (2018) 037, [\href{http://arxiv.org/abs/1901.05289}{{\tt arXiv:1901.05289}}].

\bibitem{Bhattacharyya:2018fwb}
A.~Bhattacharyya, U.~Alam, K.~L. Pandey, S.~Das, and S.~Pal, {\it {Are $H_0$
  and $\sigma_8$ tensions generic to present cosmological data?}},  {\em
  Astrophys. J.} {\bf 876} (2019), no.~2 143,
  [\href{http://arxiv.org/abs/1805.04716}{{\tt arXiv:1805.04716}}].

\bibitem{Bargiacchi:2023rfd}
G.~Bargiacchi, M.~G. Dainotti, and S.~Capozziello, {\it {Tensions with the flat
  $\boldsymbol{\Lambda}$CDM model from high-redshift cosmography}},  {\em Mon.
  Not. Roy. Astron. Soc.} {\bf 525} (2023), no.~2 3104--3116,
  [\href{http://arxiv.org/abs/2307.15359}{{\tt arXiv:2307.15359}}].

\bibitem{Haridasu:2018gqm}
B.~S. Haridasu, V.~V. Lukovi\'c, M.~Moresco, and N.~Vittorio, {\it {An improved
  model-independent assessment of the late-time cosmic expansion}},  {\em JCAP}
  {\bf 10} (2018) 015, [\href{http://arxiv.org/abs/1805.03595}{{\tt
  arXiv:1805.03595}}].

\bibitem{Bernardo:2021mfs}
R.~C. Bernardo and J.~Levi~Said, {\it {Towards a model-independent
  reconstruction approach for late-time Hubble data}},  {\em JCAP} {\bf 08}
  (2021) 027, [\href{http://arxiv.org/abs/2106.08688}{{\tt arXiv:2106.08688}}].

\bibitem{Wei:2020suh}
J.-J. Wei and F.~Melia, {\it {Cosmology-independent Estimate of the Hubble
  Constant and Spatial Curvature Using Time-delay Lenses and Quasars}},  {\em
  Astrophys. J.} {\bf 897} (2020), no.~2 127,
  [\href{http://arxiv.org/abs/2005.10422}{{\tt arXiv:2005.10422}}].

\bibitem{Naik:2023yhl}
D.~M. Naik, N.~S. Kavya, L.~Sudharani, and V.~Venkatesha, {\it
  {Model-independent cosmological insights from three newly reconstructed
  deceleration parameters with observational data}},  {\em Phys. Lett. B} {\bf
  844} (2023) 138117.

\bibitem{Capozziello:2021xjw}
S.~Capozziello, P.~K.~S. Dunsby, and O.~Luongo, {\it {Model-independent
  reconstruction of cosmological accelerated\textendash{}decelerated phase}},
  {\em Mon. Not. Roy. Astron. Soc.} {\bf 509} (2021), no.~4 5399--5415,
  [\href{http://arxiv.org/abs/2106.15579}{{\tt arXiv:2106.15579}}].

\bibitem{Alfano:2023evg}
A.~C. Alfano, C.~Cafaro, S.~Capozziello, and O.~Luongo, {\it {Dark
  energy\textendash{}matter equivalence by the evolution of cosmic equation of
  state}},  {\em Phys. Dark Univ.} {\bf 42} (2023) 101298,
  [\href{http://arxiv.org/abs/2306.08396}{{\tt arXiv:2306.08396}}].

\bibitem{Capozziello:2014zda}
S.~Capozziello, O.~Farooq, O.~Luongo, and B.~Ratra, {\it {Cosmographic bounds
  on the cosmological deceleration-acceleration transition redshift in
  $f(\mathfrak{R})$ gravity}},  {\em Phys. Rev. D} {\bf 90} (2014), no.~4
  044016, [\href{http://arxiv.org/abs/1403.1421}{{\tt arXiv:1403.1421}}].

\bibitem{Dinda:2022vmb}
B.~R. Dinda, {\it {Minimal model-dependent constraints on cosmological nuisance
  parameters and cosmic curvature from combinations of cosmological data}},
  {\em Int. J. Mod. Phys. D} {\bf 32} (2023), no.~11 2350079,
  [\href{http://arxiv.org/abs/2209.14639}{{\tt arXiv:2209.14639}}].

\bibitem{Dinda:2022jih}
B.~R. Dinda and N.~Banerjee, {\it {Model independent bounds on type Ia
  supernova absolute peak magnitude}},  {\em Phys. Rev. D} {\bf 107} (2023),
  no.~6 063513, [\href{http://arxiv.org/abs/2208.14740}{{\tt
  arXiv:2208.14740}}].

\bibitem{Dinda:2019mev}
B.~R. Dinda, {\it {Model independent parametrization of the late time cosmic
  acceleration: Constraints on the parameters from recent observations}},  {\em
  Phys. Rev. D} {\bf 100} (2019), no.~4 043528,
  [\href{http://arxiv.org/abs/1904.10418}{{\tt arXiv:1904.10418}}].

\bibitem{Ruiz-Zapatero:2022zpx}
J.~Ruiz-Zapatero, C.~Garc\'\i{}a-Garc\'\i{}a, D.~Alonso, P.~G. Ferreira, and
  R.~D.~P. Grumitt, {\it {Model-independent constraints on \ensuremath{\Omega}m
  and H(z) from the link between geometry and growth}},  {\em Mon. Not. Roy.
  Astron. Soc.} {\bf 512} (2022), no.~2 1967--1984,
  [\href{http://arxiv.org/abs/2201.07025}{{\tt arXiv:2201.07025}}].

\bibitem{Avila:2022xad}
F.~Avila, A.~Bernui, A.~Bonilla, and R.~C. Nunes, {\it {Inferring $S_8(z)$ and
  $\gamma (z)$ with cosmic growth rate measurements using machine learning}},
  {\em Eur. Phys. J. C} {\bf 82} (2022), no.~7 594,
  [\href{http://arxiv.org/abs/2201.07829}{{\tt arXiv:2201.07829}}].

\bibitem{LHuillier:2017ani}
B.~L'Huillier, A.~Shafieloo, and H.~Kim, {\it {Model-independent cosmological
  constraints from growth and expansion}},  {\em Mon. Not. Roy. Astron. Soc.}
  {\bf 476} (2018), no.~3 3263--3268,
  [\href{http://arxiv.org/abs/1712.04865}{{\tt arXiv:1712.04865}}].

\bibitem{Lee:2013sya}
S.~Lee, {\it {Measuring the matter energy density and Hubble parameter from
  Large Scale Structure}},  {\em JCAP} {\bf 02} (2014) 021,
  [\href{http://arxiv.org/abs/1307.6619}{{\tt arXiv:1307.6619}}].

\bibitem{Holanda:2019sod}
R.~F.~L. Holanda, R.~S. Gon\c{c}alves, J.~E. Gonzalez, and J.~S. Alcaniz, {\it
  {An estimate of the dark matter density from galaxy clusters and supernovae
  data}},  {\em JCAP} {\bf 11} (2019) 032,
  [\href{http://arxiv.org/abs/1905.09689}{{\tt arXiv:1905.09689}}].

\bibitem{williams1995gaussian}
C.~Williams and C.~Rasmussen, {\it Gaussian processes for regression},  {\em
  Advances in neural information processing systems} {\bf 8} (1995).

\bibitem{GpRasWil}
C.~E. Rasmussen and C.~K.~I. Williams, {\em Gaussian Processes for Machine
  Learning}.
\newblock The MIT Press, second~ed., 2006.

\bibitem{Seikel_2012}
M.~Seikel, C.~Clarkson, and M.~Smith, {\it Reconstruction of dark energy and
  expansion dynamics using gaussian processes},  {\em Journal of Cosmology and
  Astroparticle Physics} {\bf 2012} (jun, 2012) 036--036.

\bibitem{PhysRevD.85.123530}
A.~Shafieloo, A.~G. Kim, and E.~V. Linder, {\it Gaussian process cosmography},
  {\em Phys. Rev. D} {\bf 85} (Jun, 2012) 123530.

\bibitem{Hwang:2022hla}
S.-g. Hwang, B.~L'Huillier, R.~E. Keeley, M.~J. Jee, and A.~Shafieloo, {\it
  {How to use GP: Effects of the mean function and hyperparameter selection on
  Gaussian Process regression}},  \href{http://arxiv.org/abs/2206.15081}{{\tt
  arXiv:2206.15081}}.

\bibitem{Keeley:2020aym}
R.~E. Keeley, A.~Shafieloo, G.-B. Zhao, J.~A. Vazquez, and H.~Koo, {\it
  {Reconstructing the Universe: Testing the Mutual Consistency of the Pantheon
  and SDSS/eBOSS BAO Data Sets with Gaussian Processes}},  {\em Astron. J.}
  {\bf 161} (2021), no.~3 151, [\href{http://arxiv.org/abs/2010.03234}{{\tt
  arXiv:2010.03234}}].

\bibitem{Perenon:2022fgw}
L.~Perenon, M.~Martinelli, R.~Maartens, S.~Camera, and C.~Clarkson, {\it
  {Measuring dark energy with expansion and growth}},  {\em Phys. Dark Univ.}
  {\bf 37} (2022) 101119, [\href{http://arxiv.org/abs/2206.12375}{{\tt
  arXiv:2206.12375}}].

\bibitem{Perenon:2021uom}
L.~Perenon, M.~Martinelli, S.~Ili\'c, R.~Maartens, M.~Lochner, and C.~Clarkson,
  {\it {Multi-tasking the growth of cosmological structures}},  {\em Phys. Dark
  Univ.} {\bf 34} (2021) 100898, [\href{http://arxiv.org/abs/2105.01613}{{\tt
  arXiv:2105.01613}}].

\bibitem{OColgain:2021pyh}
E.~\'O~Colg\'ain and M.~M. Sheikh-Jabbari, {\it {Elucidating cosmological model
  dependence with $H_0$}},  {\em Eur. Phys. J. C} {\bf 81} (2021), no.~10 892,
  [\href{http://arxiv.org/abs/2101.08565}{{\tt arXiv:2101.08565}}].

\bibitem{Banerjee:2023evd}
N.~Banerjee, P.~Mukherjee, and D.~Pav\'on, {\it {Checking the second law at
  cosmic scales}},  {\em JCAP} {\bf 11} (2023) 092,
  [\href{http://arxiv.org/abs/2309.12298}{{\tt arXiv:2309.12298}}].

\bibitem{Banerjee:2023rvg}
N.~Banerjee, P.~Mukherjee, and D.~Pav\'on, {\it {Spatial curvature and
  thermodynamics}},  {\em Mon. Not. Roy. Astron. Soc.} {\bf 521} (2023), no.~4
  5473--5482, [\href{http://arxiv.org/abs/2301.09823}{{\tt arXiv:2301.09823}}].

\bibitem{Mukherjee:2022ujw}
P.~Mukherjee and N.~Banerjee, {\it {Constraining the curvature density
  parameter in cosmology}},  {\em Phys. Rev. D} {\bf 105} (2022), no.~6 063516,
  [\href{http://arxiv.org/abs/2202.07886}{{\tt arXiv:2202.07886}}].

\bibitem{Mukherjee:2020vkx}
P.~Mukherjee and N.~Banerjee, {\it {Revisiting a non-parametric reconstruction
  of the deceleration parameter from combined background and the growth rate
  data}},  {\em Phys. Dark Univ.} {\bf 36} (2022) 100998,
  [\href{http://arxiv.org/abs/2007.15941}{{\tt arXiv:2007.15941}}].

\bibitem{Mukherjee:2020ytg}
P.~Mukherjee and N.~Banerjee, {\it {Non-parametric reconstruction of the
  cosmological $jerk$ parameter}},  {\em Eur. Phys. J. C} {\bf 81} (2021),
  no.~1 36, [\href{http://arxiv.org/abs/2007.10124}{{\tt arXiv:2007.10124}}].

\bibitem{Li:2019nux}
E.-K. Li, M.~Du, Z.-H. Zhou, H.~Zhang, and L.~Xu, {\it {Testing the effect of
  $H_0$ on $f\sigma_8$ tension using a Gaussian process method}},  {\em Mon.
  Not. Roy. Astron. Soc.} {\bf 501} (2021), no.~3 4452--4463,
  [\href{http://arxiv.org/abs/1911.12076}{{\tt arXiv:1911.12076}}].

\bibitem{Ruiz-Zapatero:2022xbv}
J.~Ruiz-Zapatero, D.~Alonso, P.~G. Ferreira, and C.~Garcia-Garcia, {\it {Impact
  of the Universe\textquoteright{}s expansion rate on constraints on modified
  growth of structure}},  {\em Phys. Rev. D} {\bf 106} (2022), no.~8 083523,
  [\href{http://arxiv.org/abs/2207.09896}{{\tt arXiv:2207.09896}}].

\bibitem{Sakr:2023hrl}
Z.~Sakr, {\it {Testing the hypothesis of a matter density discrepancy within
  LCDM model using multiple probes}},  {\em Phys. Rev. D} {\bf 108} (2023),
  no.~8 083519, [\href{http://arxiv.org/abs/2305.02846}{{\tt
  arXiv:2305.02846}}].

\bibitem{Dinda:2023mad}
B.~R. Dinda and N.~Banerjee, {\it {Constraints on the speed of sound in the
  k-essence model of dark energy}},
  \href{http://arxiv.org/abs/2309.10538}{{\tt arXiv:2309.10538}}.

\bibitem{Dinda:2018ojk}
B.~R. Dinda, {\it {Nonlinear power spectrum in clustering and smooth dark
  energy models beyond the BAO scale}},  {\em J. Astrophys. Astron.} {\bf 40}
  (2019), no.~2 12, [\href{http://arxiv.org/abs/1804.07953}{{\tt
  arXiv:1804.07953}}].

\bibitem{Moresco:2020fbm}
M.~Moresco, R.~Jimenez, L.~Verde, A.~Cimatti, and L.~Pozzetti, {\it {Setting
  the Stage for Cosmic Chronometers. II. Impact of Stellar Population Synthesis
  Models Systematics and Full Covariance Matrix}},  {\em Astrophys. J.} {\bf
  898} (2020), no.~1 82, [\href{http://arxiv.org/abs/2003.07362}{{\tt
  arXiv:2003.07362}}].

\bibitem{moresco2012improved}
M.~Moresco, A.~Cimatti, R.~Jimenez, L.~Pozzetti, G.~Zamorani, M.~Bolzonella,
  J.~Dunlop, F.~Lamareille, M.~Mignoli, H.~Pearce, et~al., {\it Improved
  constraints on the expansion rate of the universe up to z~ 1.1 from the
  spectroscopic evolution of cosmic chronometers},  {\em Journal of Cosmology
  and Astroparticle Physics} {\bf 2012} (2012), no.~08 006--006.

\bibitem{Moresco:2015cya}
M.~Moresco, {\it {Raising the bar: new constraints on the Hubble parameter with
  cosmic chronometers at z \ensuremath{\sim} 2}},  {\em Mon. Not. Roy. Astron.
  Soc.} {\bf 450} (2015), no.~1 L16--L20,
  [\href{http://arxiv.org/abs/1503.01116}{{\tt arXiv:1503.01116}}].

\bibitem{Moresco:2016mzx}
M.~Moresco, L.~Pozzetti, A.~Cimatti, R.~Jimenez, C.~Maraston, L.~Verde,
  D.~Thomas, A.~Citro, R.~Tojeiro, and D.~Wilkinson, {\it {A 6\% measurement of
  the Hubble parameter at $z\sim0.45$: direct evidence of the epoch of cosmic
  re-acceleration}},  {\em JCAP} {\bf 05} (2016) 014,
  [\href{http://arxiv.org/abs/1601.01701}{{\tt arXiv:1601.01701}}].

\bibitem{zhang2014four}
C.~Zhang, H.~Zhang, S.~Yuan, S.~Liu, T.-J. Zhang, and Y.-C. Sun, {\it Four new
  observational h (z) data from luminous red galaxies in the sloan digital sky
  survey data release seven},  {\em Research in Astronomy and Astrophysics}
  {\bf 14} (2014), no.~10 1221.

\bibitem{Simon:2004tf}
J.~Simon, L.~Verde, and R.~Jimenez, {\it {Constraints on the redshift
  dependence of the dark energy potential}},  {\em Phys. Rev. D} {\bf 71}
  (2005) 123001, [\href{http://arxiv.org/abs/astro-ph/0412269}{{\tt
  astro-ph/0412269}}].

\bibitem{Ratsimbazafy:2017vga}
A.~L. Ratsimbazafy, S.~I. Loubser, S.~M. Crawford, C.~M. Cress, B.~A. Bassett,
  R.~C. Nichol, and P.~V\"ais\"anen, {\it {Age-dating Luminous Red Galaxies
  observed with the Southern African Large Telescope}},  {\em Mon. Not. Roy.
  Astron. Soc.} {\bf 467} (2017), no.~3 3239--3254,
  [\href{http://arxiv.org/abs/1702.00418}{{\tt arXiv:1702.00418}}].

\bibitem{stern2010cosmic}
D.~Stern, R.~Jimenez, L.~Verde, M.~Kamionkowski, and S.~A. Stanford, {\it
  Cosmic chronometers: constraining the equation of state of dark energy. i: H
  (z) measurements},  {\em Journal of Cosmology and Astroparticle Physics} {\bf
  2010} (2010), no.~02 008.

\bibitem{Borghi:2021rft}
N.~Borghi, M.~Moresco, and A.~Cimatti, {\it {Toward a Better Understanding of
  Cosmic Chronometers: A New Measurement of H(z) at z \ensuremath{\sim} 0.7}},
  {\em Astrophys. J. Lett.} {\bf 928} (2022), no.~1 L4,
  [\href{http://arxiv.org/abs/2110.04304}{{\tt arXiv:2110.04304}}].

\bibitem{Avila:2021dqv}
F.~Avila, A.~Bernui, E.~de~Carvalho, and C.~P. Novaes, {\it {The growth rate of
  cosmic structures in the local Universe with the ALFALFA survey}},  {\em Mon.
  Not. Roy. Astron. Soc.} {\bf 505} (2021), no.~3 3404--3413,
  [\href{http://arxiv.org/abs/2105.10583}{{\tt arXiv:2105.10583}}].

\bibitem{Hawkins:2002sg}
E.~Hawkins et~al., {\it {The 2dF Galaxy Redshift Survey: Correlation functions,
  peculiar velocities and the matter density of the universe}},  {\em Mon. Not.
  Roy. Astron. Soc.} {\bf 346} (2003) 78,
  [\href{http://arxiv.org/abs/astro-ph/0212375}{{\tt astro-ph/0212375}}].

\bibitem{Guzzo:2008ac}
L.~Guzzo et~al., {\it {A test of the nature of cosmic acceleration using galaxy
  redshift distortions}},  {\em Nature} {\bf 451} (2008) 541--545,
  [\href{http://arxiv.org/abs/0802.1944}{{\tt arXiv:0802.1944}}].

\bibitem{Blake:2013nif}
C.~Blake et~al., {\it {Galaxy And Mass Assembly (GAMA): improved cosmic growth
  measurements using multiple tracers of large-scale structure}},  {\em Mon.
  Not. Roy. Astron. Soc.} {\bf 436} (2013) 3089,
  [\href{http://arxiv.org/abs/1309.5556}{{\tt arXiv:1309.5556}}].

\bibitem{Blake_2011}
C.~Blake, S.~Brough, M.~Colless, C.~Contreras, W.~Couch, S.~Croom, T.~Davis,
  M.~J. Drinkwater, K.~Forster, D.~Gilbank, M.~Gladders, K.~Glazebrook,
  B.~Jelliffe, R.~J. Jurek, I.-h. Li, B.~Madore, D.~C. Martin, K.~Pimbblet,
  G.~B. Poole, M.~Pracy, R.~Sharp, E.~Wisnioski, D.~Woods, T.~K. Wyder, and
  H.~K.~C. Yee, {\it The wigglez dark energy survey: the growth rate of cosmic
  structure since redshift z=0.9: Wigglez survey: growth of structure},  {\em
  Monthly Notices of the Royal Astronomical Society} {\bf 415} (June, 2011)
  2876–2891.

\bibitem{SDSS:2006lmn}
{\bf SDSS} Collaboration, M.~Tegmark et~al., {\it {Cosmological Constraints
  from the SDSS Luminous Red Galaxies}},  {\em Phys. Rev. D} {\bf 74} (2006)
  123507, [\href{http://arxiv.org/abs/astro-ph/0608632}{{\tt
  astro-ph/0608632}}].

\bibitem{Ross:2006me}
N.~P. Ross et~al., {\it {The 2dF-SDSS LRG and QSO Survey: The 2-Point
  Correlation Function and Redshift-Space Distortions}},  {\em Mon. Not. Roy.
  Astron. Soc.} {\bf 381} (2007) 573--588,
  [\href{http://arxiv.org/abs/astro-ph/0612400}{{\tt astro-ph/0612400}}].

\bibitem{daAngela:2006mf}
J.~da~Angela et~al., {\it {The 2dF-SDSS LRG and QSO Survey: QSO clustering and
  the L-z degeneracy}},  {\em Mon. Not. Roy. Astron. Soc.} {\bf 383} (2008)
  565--580, [\href{http://arxiv.org/abs/astro-ph/0612401}{{\tt
  astro-ph/0612401}}].

\bibitem{Cao:2019kgn}
S.~Cao, J.~Qi, Z.~Cao, M.~Biesiada, J.~Li, Y.~Pan, and Z.-H. Zhu, {\it {Direct
  test of the FLRW metric from strongly lensed gravitational wave
  observations}},  {\em Sci. Rep.} {\bf 9} (2019), no.~1 11608,
  [\href{http://arxiv.org/abs/1910.10365}{{\tt arXiv:1910.10365}}].

\bibitem{Liu:2020pfa}
Y.~Liu, S.~Cao, T.~Liu, X.~Li, S.~Geng, Y.~Lian, and W.~Guo, {\it
  {Model-independent constraints on cosmic curvature: implication from updated
  Hubble diagram of high-redshift standard candles}},  {\em Astrophys. J.} {\bf
  901} (2020), no.~2 129, [\href{http://arxiv.org/abs/2008.08378}{{\tt
  arXiv:2008.08378}}].

\bibitem{Freedman:2019jwv}
W.~L. Freedman et~al., {\it {The Carnegie-Chicago Hubble Program. VIII. An
  Independent Determination of the Hubble Constant Based on the Tip of the Red
  Giant Branch}},  {\em Astrophys. J.} {\bf 882} (2019) 34,
  [\href{http://arxiv.org/abs/1907.05922}{{\tt arXiv:1907.05922}}].

\bibitem{Riess:2020fzl}
A.~G. Riess, S.~Casertano, W.~Yuan, J.~B. Bowers, L.~Macri, J.~C. Zinn, and
  D.~Scolnic, {\it {Cosmic Distances Calibrated to 1\% Precision with Gaia EDR3
  Parallaxes and Hubble Space Telescope Photometry of 75 Milky Way Cepheids
  Confirm Tension with $\Lambda$CDM}},  {\em Astrophys. J. Lett.} {\bf 908}
  (2021), no.~1 L6, [\href{http://arxiv.org/abs/2012.08534}{{\tt
  arXiv:2012.08534}}].

\end{thebibliography}\endgroup

\end{document}